\documentclass[aps,preprintnumbers,onecolumn,amsmath,amssymb,floatfix,pra,notitlepage]{revtex4-1} 
\pdfoutput=1
\usepackage[utf8]{inputenc}
\usepackage{amsmath}
\usepackage{amsfonts}
\usepackage{braket}
\usepackage{tensor}
\usepackage{bbold}
\usepackage{comment}
\usepackage{graphicx}
\usepackage{empheq}
\usepackage[colorlinks=true,linkcolor=blue, citecolor=blue, bookmarks]{hyperref}
\def\be{\begin{equation}}
\def\ee{\end{equation}}
\def\bea{\begin{eqnarray}}
\def\eea{\end{eqnarray}}

\numberwithin{equation}{section}

\begin{document}

\title{R\'enyi entropies of generic thermodynamic macrostates in integrable systems}
\author{M\'arton Mesty\'an$^1$, Vincenzo Alba$^1$, and Pasquale Calabrese$^{1,2}$}
\affiliation{$^1$SISSA \& INFN, via Bonomea 265, 34136 Trieste, Italy}
\affiliation{$^2$International Centre for Theoretical Physics (ICTP), 34151, Trieste, Italy}
\begin{abstract}
We study the behaviour of R\'enyi entropies in a generic thermodynamic macrostate of an integrable model. 
In the standard quench action approach to quench dynamics, the R\'enyi entropies may be derived from the 
overlaps of the initial state with Bethe eigenstates. 
These overlaps fix the driving term in the thermodynamic Bethe ansatz (TBA) formalism. 
We show that this driving term can be also reconstructed starting from the macrostate's particle densities.
We then compute explicitly the stationary R\'enyi entropies after the quench from the dimer and the tilted N\'eel state in XXZ spin chains.
For the former state we employ the overlap TBA approach, while for the latter we reconstruct the driving terms from the macrostate.
We discuss in full details the limits that can be analytically handled and we use numerical simulations to check our results against the
large time limit of the entanglement entropies.  

\end{abstract}

\maketitle

\section{Introduction}
\label{sec:introduction}

The time evolution of the entanglement entropy plays a crucial role in the understanding of the non-equilibrium dynamics 
of isolated quantum systems, in particular for quantum quenches in many body systems.   
Indeed, the growth of the entanglement entropy in time has been related to the efficiency of tensor network 
algorithms \cite{swvc-08,pv-08,rev0,d-17} such as the time dependent density matrix renormalisation group (tDMRG). 
Furthermore, the extensive value (in subsystem size) reached by the entanglement entropy at long time has been understood as
the thermodynamic entropy of the ensemble 
describing stationary local properties of the system \cite{dls-13,collura-2014,bam-15,kaufman-2016,alba-2016,nahum-17,alba-2018,nwfs-18} 
and related to other thermodynamic entropy definitions \cite{collura-2014,polkovnikov-2011,santos-2011,spr-12,gurarie-2013,fagotti-2013b,kormos-2014,Rigo14,Rigo16,piroli-2016}.

In a quantum quench~\cite{calabrese-2006,gogolin-2015,calabrese-2016,ef-16}, a many-body system is initially prepared in a low-entanglement state 
$|\Psi_0\rangle$ and is let evolve with a many-body Hamiltonian $H$ such that $[H,|\Psi_0\rangle\langle \Psi_0|]\neq 0$. 
In this protocol, the entire system is in a pure state at any time,  but the reduced density matrix of an 
arbitrary finite compact subsystem attains a long time limit that can be described by a statistical ensemble (see, e.g., Ref. \cite{ef-16}).
Thus, at asymptotically long times, all {\it local} physical observables relax to stationary values.
The properties of the reduced density matrix are captured by a Gibbs (thermal) ensemble
for generic systems~\cite{eth0,deutsch-1991,srednicki-1994,rdo-08,rigol-2012,dalessio-2015}
and by a generalised Gibbs ensemble (GGE)  for integrable systems~
\cite{rigol-2007,bastianello-2016,vernier-2016,alba-2015,cc-07,
cramer-2008,barthel-2008,cramer-2010,calabrese-2011,calabrese-2012,cazalilla-2006,cazalilla-2012a,
sotiriadis-2012,collura-2013,fagotti-2013,kcc14,sotiriadis-2014,essler-2015,vidmar-2016,pvw-17,
ilievski-2015a,cardy-2015,sotiriadis-2016,mc-12,p-13,fe-13b,fcec-14,langen-15,pk-18}.

In one-dimensional systems, the entanglement entropy after a global quantum quench has 
been found \cite{calabrese-2005,fagotti-2008,alba-2016,dmcf-06,ep-08,lauchli-2008,kim-2013,nr-14,coser-2014,cce-15,fc-15,cc-16,chmm-16,buyskikh-2016,kctc-17,mkz-17,p-18,fnr-17,bhy-17,hbmr-17,nahum-17,alba-2016,alba-2018,r-2017,BeTC18,mbpc-17,nrv-18}
to generically grow linearly in time up to a point (linear in subsystem size) when a saturation regime to an extensive value takes place. 
This time evolution of the von Neumann entanglement entropy for a generic integrable system may be fully understood in terms of 
a semiclassical quasiparticle picture \cite{calabrese-2005} for the spreading of the entanglement, complemented 
with the Bethe ansatz knowledge of the stationary state, as shown in \cite{alba-2016}. 
These ideas have been first developed for the time evolution after a quantum quench in integrable models \cite{alba-2016}, 
but later they have been generalised to other situations as, e.g., in Refs. \cite{dsvc17,kz-17,alba-inh,ckt-18,bfpc-18,eb-17,ctd-18}. 
However, in spite of its great success, this approach hardly generalises to other entanglement estimators and in particular 
to the R\'enyi entanglement entropies  defined as 
\begin{equation}
\label{sr}
S^{(\alpha)}[\rho_A]\equiv\frac{1}{1-\alpha}\ln\textrm{Tr}\rho_A^\alpha,
\end{equation}
in terms of the RDM $\rho_A$ of the considered subsystem $A$.  
The R\'enyi entropies are very important physical quantities for several reasons. 
First, their knowledge for different values of the index $\alpha$ gives access to the entire spectrum of the density 
matrix (see e.g. \cite{calabrese-2008}). 
Since in the limit $\alpha\to1$ one has $S^{(\alpha)}[\rho_A]\to  S[\rho_A]\equiv -{\rm Tr}\rho_A\ln \rho_A$, 
they represent the essence of the replica approach to the entanglement entropy \cite{cc-04}. 
While the replica method was originally introduced as a theoretical analytic tool to deal with the complexity 
of $\rho_A$ \cite{cc-04}, it has become 
a fundamental concept to access the entanglement entropy in numerical approaches based on Monte Carlo simulations \cite{mc}
and also in real experiments \cite{daley-2012,evdcz-18}: R\'enyi entanglement entropies (for $\alpha=2$) have been measured 
experimentally with cold atoms, both in equilibrium~\cite{islam-2015} and after a quantum quench~\cite{kaufman-2016}. 
Unfortunately,  only for non-interacting systems we know how to generalise the quasiparticle picture for the spreading 
of the entanglement to the R\'enyi entropies (see, e.g., \cite{p-18, AlCa17}). 
For interacting integrable models, the Thermodynamic Bethe Ansatz (TBA) approach to quantum quenches 
(overlap TBA or Quench Action method~\cite{caux-2013,caux-16}) has been adapted in \cite{AlCa17} 
to the calculation of the stationary value of the R\'enyi entropies for both the diagonal and GGE ensembles. 
Within this approach, in the thermodynamic limit, $S^{(\alpha)}$ is given as a generalised free energy over
a  saddle point eigenstate (representative eigenstate or thermodynamic macrostate) which is {\it not} the one corresponding to the 
stationary state describing local observables and von Neumann entropy, as we shall review later. 
It turns out that this shifting of the saddle point is the main obstacle toward a quantitative  semiclassical formula for the time evolution 
of the R\'enyi entanglement entropies whose large time limit is provided by the GGE value. 

This general approach, so far has been used for the calculation of the R\'enyi entropies only for the quench from the N\'eel state in the 
XXZ spin-chain \cite{AlCa17,AlCa17b}, finding a very interesting $\alpha$ and $\Delta$ dependence. 
The goal of this manuscript is twofold. On the one hand, we will apply the approach of Refs. \cite{AlCa17,AlCa17b} to 
the calculation of the stationary R\'enyi entropies after the quench from other initial states, a problem {\it per se} of high interest. 
On the other hand, this approach strongly depends on the knowledge of the overlaps between the initial state and the 
Bethe states. These are known in many cases, but the GGE can be also constructed in a simpler way from the conservation 
of all local and quasilocal charges \cite{ilievski-2015a,pvw-17}. 
In particular until very recently, for many initial states (i.e. tilted N\'eel and tilted ferromagnet) 
in the quench in XXZ spin-chain, stationary values were known only by means of the latter method \cite{pvc-16}. 
For this reason, we developed a hybrid numerical/analytic method to get exact predictions for the R\'enyi entropies which 
does not require an a-priori knowledge of the overlaps and apply it to the prototypical case of tilted N\'eel states. 
However, after the completion of our calculations, a manuscript appeared \cite{Pozsgay18} providing  a conjecture 
for the overlaps also in this case. 
These served as a test of the validity of our method and did not alter the logic of the calculation which was
a proof of concepts about the construction of R\'enyi entropies without knowing the overlaps.

The paper is organised as follows.
In Sec. \ref{sec:xxzModel} we introduce the XXZ spin-chain and review its Bethe ansatz solution. 
In Sec. \ref{sec:renyiEntropyTBA} we first review the TBA approach for the R\'enyi entropies \cite{AlCa17},
work out some major  simplifications on the known expressions, and finally we apply this machinery 
to the quench from the dimer state, working out explicitly a two limits ($\Delta\to\infty$ and $\alpha\to\infty$) 
that can be handled analytically. 
In Sec. \ref{sec:sourceTermExtraction} we show how to calculate the R\'enyi entropies for generic macrostates 
and we apply this machinery to the explicit case of the tilted N\'eel state, finding several interesting new results.
In Sec. \ref{sec:dmrg} we test, by means of extensive tDMRG simulations, 
that the thermodynamic R\'enyi entropies agree with the long time limit of the 
entanglement entropies. 
Finally in Sec. \ref{sec:conclusions} we summarise our findings and discuss open problems.

\section{The XXZ model and its Bethe ansatz solution}
\label{sec:xxzModel}

In this work we consider quantum quenches in the spin-$1/2$ one-dimensional 
anisotropic Heisenberg model  ($XXZ$ spin chain) with Hamiltonian 
\begin{equation}
  H = \sum_{k=1}^{L} \left[ \sigma_{k}^{x}\sigma_{k+1}^{x} + \sigma_{k}^{y}\sigma_{k+1}^{y} + \Delta\,\left(\sigma_{k}^{z}\sigma_{k+1}^{z}-1\right)\right], 
  \label{eq:hXXZHamiltonian}
\end{equation}
where $\sigma^{x,y,z}_{j}$ are the Pauli matrices and $\Delta$ is the anisotropy 
parameter. We focus on the antiferromagnetic gapped phase for $\Delta > 1$. Periodic boundary 
conditions are imposed by choosing $\sigma^{x,y,z}_{L+1}=\sigma^{x,y,z}_{1}$.
The Hamiltonian of the XXZ chain commutes with the total magnetisation $S_T^z\equiv 
\sum_i\sigma_i^z/2$. As a consequence, the eigenstates of~\eqref{eq:hXXZHamiltonian} can 
be labelled by the value of $S_T^z$. Due to periodic boundary conditions, the XXZ chain is 
invariant under one site translations $\sigma_i^\alpha\to\sigma_{i+1}^\alpha$. This 
means that $[{\mathcal T},H]=0$, where ${\mathcal T}$ is the one-site translation operator. 

In the following, we will consider quantum quenches from the Majumdar-Ghosh (dimer) 
state and from the tilted N\'eel state. The dimer state is defined as 
\begin{equation}
  |\mathrm D \rangle = 
  	\left( 
  	\frac{|\uparrow\downarrow\rangle - |\downarrow\uparrow\rangle}{\sqrt{2}}
	\right)^{\otimes L/2}.
  \label{eq:dimerInitialState}
\end{equation}
To take advantage of the translational invariance of the XXZ chain, 
we will consider the translation invariant version of the dimer state, which 
is given as 
 \begin{equation}
   |\Psi_{0} \rangle = \left( \frac{1 + \mathcal{T}}{\sqrt{2}} \right)\, | \mathrm D \rangle,
   \label{eq:pTranslationallyInvariantDimerState}
 \end{equation}
where $ \mathcal{T}$ is the one-site translation operator. 
Indeed, the stationary states for the quench from the dimer state \eqref{eq:dimerInitialState} and from its translational invariant 
version \eqref{eq:pTranslationallyInvariantDimerState} are the same, but the time evolution is clearly different \cite{fcec-14}. 

We also consider quenches from the tilted N\'eel state defined as 
\begin{equation}
  |\Psi_{0} \rangle = \left[  \left( \cos(\theta/2) |\!  \uparrow \rangle + i \sin (\theta/2)  |\! \downarrow \rangle \right) \otimes \left( -i \sin (\theta/2) |\! \uparrow \rangle +  \cos(\theta/2) |\! \downarrow \rangle \right)  \right]^{\otimes L/2}.
  \label{eq:pTiltedNeelInitialState}
\end{equation}
Here $\theta$ is the tilting angle. The tilted N\'eel state is obtained by 
applying the global rotation operator to the translational invariant version 
of the N\'eel state $|N\rangle\equiv|\uparrow\downarrow\cdots\rangle$ as 
\begin{equation}
|\Psi_0\rangle=e^{i\theta/2\sum_i\sigma_i^y}
\Big(\frac{|N\rangle+{\mathcal T}|N\rangle}{\sqrt{2}}\Big).
\end{equation}
%

In this work we are interested in the steady-state R\'enyi entropies after 
the quenches from~\eqref{eq:pTranslationallyInvariantDimerState} 
and~\eqref{eq:pTiltedNeelInitialState}. 
For the dimer state we proceed using the technique developed in Refs~\cite{AlCa17,AlCa17b}, 
i.e. exploiting the analytic knowledge of  the overlaps between the state and the eigenstates of the XXZ chain.
By using the Quench Action method~\cite{caux-2013,caux-16} this provides  
a set of driving functions for a generalised set of TBA equations that determine the steady-state R\'enyi entropies. 
Conversely, when we started this work, the overlaps between the tilted N\'eel state and the 
eigenstates of the XXZ chain were not known (see Ref.~\cite{Pozsgay18} for a recent conjecture).
Thus we provide a suitable generalisation of the method of Ref. \cite{AlCa17} to reconstruct the driving terms from 
the stationary ensemble.
This generalisation is one of the main technical results of the paper.

\subsection{Bethe ansatz solution of the XXZ chain}
\label{sec-BA}

The XXZ chain can be solved by the Bethe ansatz~\cite{Taka99}, which allows one 
to construct {\it all} the eigenstates of~\eqref{eq:hXXZHamiltonian} analytically. 
It is convenient to work in the sector with fixed magnetisation $S_T^z$, or, equivalently 
with fixed number $N$ of down spins. Here we follow  the standard Bethe ansatz 
framework referring to down spins as particles. The eigenstates of the 
XXZ chain are easily constructed 
starting from the ferromagnetic state, i.e., the state with all the spins up 
$|\uparrow\uparrow\cdots\rangle$. The Bethe ansatz expression for the generic 
eigenstate in the sector with $N$ down spins reads  
\begin{equation}
| \boldsymbol \lambda \rangle =  \sum_{n_{1}< n_{2} < \dots < n_{N}} \psi(n_{1},n_{2},\dots,n_{N}) \sigma^{-}_{n_{1}} \sigma^{-}_{n_{2}} \dots \sigma^{-}_{n_{N}} |\! \uparrow \uparrow\dots \uparrow \rangle, 
  \label{eq:pBetheAnsatzEigenstate1} 
\end{equation}
where $\sigma_i^-$ is the spin-$1/2$ lowering operator. 
Here the sum is over the positions $n_i$ of the particles. The 
amplitudes $\psi(n_1,n_2,\dots,n_N)$ read 
\begin{equation}
 \label{eq:pBetheAnsatzEigenstate2} 
\psi(n_{1},n_{2},\dots,n_{n}) = \sum_{P} A(P) \prod_{j=1}^{N} \left(\frac{\sin(\lambda_{P_j}+i \eta/2)}{\sin(\lambda_{P_j} - i\eta/2 )}\right)^{n_{j}},
\end{equation}
where $\eta\equiv\textrm{arcosh}(\Delta)$, and the sum is now over the permutations 
$P$ of the indices $j\in[1,N]$. The overall amplitude $A(P)$ is given as  
\begin{equation}
A(P) = (-1)^{\textrm{sign}(P)} \prod_{j=1}^{N} \prod_{k=j+1}^{N} \sin(\lambda_{P_{j}}-\lambda_{P_{k}} + i\eta).
\label{eq:pBetheAnsatzEigenstate3} 
\end{equation}
The state is then specified by a set of $N$ rapidities $\lambda_j$ that play the same role of 
the quasimomenta in free models. 
The total energy of the eigenstate is obtained by summing independently the 
contributions of each particle, to obtain 
\begin{equation}
 E = \sum_{j=1}^{N} e(\lambda_{j}),\qquad   e(\lambda) \equiv - \frac{4 \sinh(\eta)^{2}}{\cosh(\eta)-\cos(2\lambda_{j})}.
  \label{eq:eEigenEnergy}
\end{equation}
In order for~\eqref{eq:pBetheAnsatzEigenstate1} to be an eigenstate of the XXZ 
chain, the rapidities have to be solutions of the Bethe equations~\cite{Taka99} 
\begin{equation}
  \left( \frac{\lambda_{j} + i\eta/2}{\lambda_{j} - i\eta/2} \right)^{L} 
  =
  - \prod_{k=1}^{N} \frac{\lambda_{j} - \lambda_{k} + i \eta}{\lambda_{j} - \lambda_{k} - i\eta}
  \qquad
  (j=1,\dots,N).
  \label{eq:lBetheEquationsFiniteSize}
\end{equation}
Each independent set of solutions $\{\lambda_j\}_{j=1}^N$ identifies a 
different eigenstate of the XXZ chain. 

A distinctive property of the XXZ model, which underlies its integrability, 
is the existence of an infinite number of pairwise commuting {\it local} and {\it quasilocal} 
conserved quantities (charges)~\cite{ilievski-2016}
\begin{equation}
  Q_{s}^{(j)} = \frac{d^{j-1}}{d \lambda^{j-1} } T_s(\lambda) \bigg|_{\lambda=0}, \qquad  [Q_{r}^{(j)},Q_{s}^{(k)}] = 0, \qquad Q_{1/2}^{(2)} = \frac{ 1}{2 \sinh \eta} H_{\mathrm{XXZ}}.
  \label{eq:commutingConservedCharges}
\end{equation}
 Here $T_{s}(\lambda)$ is the transfer matrix of the XXZ model with an $s$-dimensional auxiliary space. The dimension $s$ of the auxiliary space can be any positive integer or half-integer. In fact, choosing $s=1/2$ yields the usual transfer matrix of the six-vertex model \cite{KoBI93}.  The corresponding charges $Q_{1/2}^{(j)}$ are local, which means they are of the sum of operators that act non-trivially at most at $m^{(j)}$ sites, where $m^{(j)}$ is an integer depending on $j$. For example, the charge $Q_{1/2}^{(2)}$ has $m^{(2)}=2$ and it is proportional to the system Hamiltonian~\eqref{eq:hXXZHamiltonian}. On the other hand, there is no such limit on the size of the terms in the charges $Q_{s}^{(j)}$ when $s>1/2$. These charges, called {\it quasilocal charges}, are crucial to correctly describe the steady state arising after a quantum quench in integrable models~\cite{ilievski-2015a}. 

Since the charges \eqref{eq:commutingConservedCharges} commute with the Hamiltonian \eqref{eq:hXXZHamiltonian}, all of them are diagonal in the basis of Bethe ansatz eigenstates \eqref{eq:pBetheAnsatzEigenstate1}-\eqref{eq:lBetheEquationsFiniteSize}. The corresponding eigenvalues are given as \cite{ilievski-2016} 
\begin{align}
  Q^{(j)}_{s}  | \boldsymbol \lambda \rangle = \sum_{k=1}^{N} q_{s}^{(j)}  (\lambda_{k})| \boldsymbol \lambda \rangle + o(L), \quad q_{s}^{(j)}(\lambda) = \left( -i \frac{d}{d \mu} \right)^{j-1} \ln  \left(\frac{\sin(\mu - \lambda + i s \eta )}{\sin(\mu - \lambda - i s \eta )}\right)\bigg|_{\mu=0},
  \label{eq:conservedChargeEigenvalues}
\end{align}
 where $o(L)$ indicates terms that are either zero (in the case of local charges with $s=1/2$) or vanish in the thermodynamic limit (in the case of quasilocal charges with $s>1/2$). 

\subsection{Thermodynamic Bethe Ansatz (TBA)}

In this paper we are interested in the thermodynamic limit $L,N\to\infty$, 
with the ratio $N/L$ (i.e., the particle density) fixed. The solutions of the 
Bethe equations~\eqref{eq:lBetheEquationsFiniteSize} are in general complex. 
However, a remarkable feature of the Bethe equations is that 
in the thermodynamic limit their solutions are organised into strings, 
i.e., multiplets of solutions having the same real part, but different 
imaginary components. This is the famous string hypothesis~\cite{Taka99}. 
Precisely, the generic structure of a string multiplet reads  
\begin{equation}
  \lambda_{n,j}^{\alpha} = \lambda_{n}^{\alpha} + \frac{i \eta}{2} (n + 1 - 2j) + \delta_{n,j}^{\alpha}.
  \label{eq:lStringRapiditiesTDL}
\end{equation}
Here $n$ is the string length, i.e., the number of solutions with the same real 
part, $\alpha$ labels the different strings of the same size $n$, and $j$ 
labels the different components within the same 
string multiplet. The {\it real} number $\lambda_{n}^{\alpha}$ is  
called string center. The string hypothesis holds 
only in the thermodynamic limit: for finite chains, string deviations 
(denoted as $\delta_{n,j}^{\alpha}$ in~\eqref{eq:lStringRapiditiesTDL}) 
are present, but for thermodynamically relevant states, they vanish exponentially in $L$. 

In the thermodynamic limit the string centers become dense 
on the real axis. Thus, instead of working with individual eigenstates, 
it is convenient to describe the thermodynamic quantities by introducing 
the densities $\rho_{n}(\lambda)$, one for each string type $n$. $\rho_n(\lambda)$ 
are the densities of string centers on the real line. 
Similarly, one can introduce the densities of holes $\rho_{\mathrm{h},n}(\lambda)$, 
which is the density of unoccupied string centers, and the total density (density of 
states) $\rho_{\mathrm{t},n}(\lambda)\equiv\rho_{n}(\lambda)+
\rho_{\mathrm{h},n}(\lambda)$. Each set of particle and hole densities identify a 
thermodynamic macrostate of the XXZ chain. 
Moreover, a given set of densities $\rho_{n}(\lambda)$ corresponds to an exponentially 
large (with $L$ and $N$) number of microscopic eigenstates. The rapidities 
identifying all these eigenstates converge in the thermodynamic limit to the 
same set of densities. The logarithm of the number of thermodynamically equivalent 
eigenstates is given by the Yang--Yang entropy~\cite{yy-69} 
\begin{equation}
S_{\mathrm{YY}}= s_{YY}L=L\sum_{n=1}^{\infty} \int\! \mathrm{d} \lambda\, \left[ \rho_{\mathrm{t}, n}(\lambda) \ln  \rho_{\mathrm{t}, n}(\lambda) - \rho_{n}(\lambda) \ln  \rho_{n}(\lambda) - \rho_{\mathrm{h}, n}(\lambda) \ln  \rho_{\mathrm{h}, n}(\lambda) \right].
  \label{eq:sYangYangEntropy}
\end{equation}
The Yang-Yang entropy is extensive, and its density is obtained by summing 
independently the contributions of the different rapidities and string 
types $n$. The only constraint that a legitimate set of densities has to 
satisfy is given by the continuum limit of the Bethe equations, which are 
called Bethe-Gaudin-Takahashi (BGT) equations~\cite{Taka99}
\begin{equation}
  \rho_{\mathrm{t}, n} = a_{n} - \sum_{m=1}^{\infty} A_{nm} \star \rho_{m},
  \label{eq:rBetheGaudinTakahashiCoupled}
\end{equation}
where we defined 
\begin{equation}
a_{n}(\lambda) \equiv \frac{1}{ \pi}  \frac{\sinh(n \eta)}{ \cosh(n \eta) - \cos (2 \lambda)}.
\label{eq:aBGTCoupledSourceAN}
\end{equation}
In~\eqref{eq:rBetheGaudinTakahashiCoupled}, $A_{nm}$ are the scattering 
phases between the bound states,  defined as 
\begin{equation}
A_{nm}(\lambda) \equiv (1-\delta_{n,m})a_{|n-m|}(\lambda) + a_{|n-m|+2}(\lambda) + \dots + a_{n+m-2}(\lambda) + a_{n+m}(\lambda),
\label{eq:aBGTCoupledKernelANM}
\end{equation}
and the star symbol $\star$ denotes the convolution 
\begin{equation}
  [f \star g](\lambda) \equiv \int_{-\pi/2}^{\pi/2} \! \mathrm{d} \mu\, f(\lambda-\mu) g(\mu).
  \label{eq:fConvolutionDefinition}
\end{equation}

Importantly, a standard trick in Bethe ansatz allows one to simplify~\eqref{eq:rBetheGaudinTakahashiCoupled}
obtaining a system of partially decoupled integral equations as~\cite{Taka99} 
\begin{equation}
  \rho_{\mathrm{t}, n} = s \star (\rho_{\mathrm{h}, n-1} + \rho_{\mathrm{h}, n+1}) \quad (n=1,2,\dots),
  \label{eq:rBetheGaudinTakahashiDecoupled}
\end{equation}
where, for the sake of simplicity, we defined $\rho_{\mathrm{h}, 0}(\lambda) 
\equiv \delta(\lambda) $, and we introduced $s(\lambda)$ as 
\begin{equation}
  s(\lambda) \equiv \frac{1}{2 \pi} \sum_{k=-\infty}^{\infty} \frac{e^{-2 i k \lambda}}{ \cosh{ k \eta}} = \frac{1}{2 \pi} + \frac{1}{ 2 \pi} \sum_{k=1}^{\infty} \frac{ \cos 2 k \lambda}{ \cosh k \eta}.
  \label{eq:sFunction}
\end{equation}
The partially decoupled system is typically easier to solve numerically 
than~\eqref{eq:rBetheGaudinTakahashiCoupled}.

In the thermodynamic limit, the eigenvalues of the conserved 
quantities $Q_s^{(j)}$~\eqref{eq:conservedChargeEigenvalues}  can be written in terms of the densities $\rho_n$ 
as 
\begin{align}
  \langle \boldsymbol \rho | Q^{(j)}_{s} | \boldsymbol \rho \rangle = \sum_{n=1}^{\infty} \int_{-\pi/2}^{\pi/2} d \lambda \rho_{n}(\lambda) q_{s,n}^{(j)}(\lambda), \qquad q_{s,n}^{(j)} = \sum_{k=1}^{n} q_{s}^{(j)}\left(\lambda + \frac{i \eta}{2} (n + 1 - 2k)\right).
  \label{eq:conservedChargeEigenvaluesTBA}
\end{align}
Here $q_s^{(j)}$ is the same as in~\eqref{eq:conservedChargeEigenvalues}. 

\section{TBA approach for the stationary R\'enyi entropies}
\label{sec:renyiEntropyTBA}

In this section we summarise the recently developed TBA approach to 
compute R\'enyi entropies in the steady state at long time after a quantum 
quench~\cite{AlCa17,AlCa17b}. 
In integrable models, the steady-state can be characterised in 
terms of the initial-state expectation value of the 
infinite set of local and quasilocal conserved charges~\eqref{eq:commutingConservedCharges}. 
These conserved quantities are key to construct the GGE that describes local 
and quasilocal observables in the steady state. The density matrix of the GGE is
\begin{equation}
  \rho_{\mathrm{GGE}} \equiv \frac{1}{Z_{\mathrm{GGE}}}\, \exp \left(- \sum_{s,j} \beta_{s}^{(j)}   Q_{s}^{(j)} \right),\qquad  Z_{\mathrm{GGE}} \equiv \mathrm{Tr }\,\exp \left(- \sum_{s,j} \beta_{s}^{(j)}  Q_{s}^{(j)} \right),
  \label{eq:rGGEDensityMatrix}
\end{equation}
where $Q_{s}^{(j)}$ are the local and quasilocal conserved charges ($s=1/2,1,3/2,2,\dots$ and $j=1,2,\dots$), 
and $Z_{\textrm{GGE}}$ is a normalisation. The Lagrange multipliers $\beta_{s}^{(j)}$ fix the GGE expectation values of the 
charges to their initial-state values $\langle \Psi_{0} | Q_{s}^{(j)} | \Psi_{0} \rangle$.
Similarly to the standard (thermal) TBA, local and 
quasilocal properties of the steady state are fully encoded in an appropriate 
thermodynamic macrostate~\cite{caux-2013,caux-16}, which is fully characterised 
by a set of densities $\rho_n$ and $\rho_{\textrm{h},n}$.

The GGE R\'enyi entropies are by definition 
\begin{equation}
S_{\mathrm{GGE}}^{(\alpha)} = \frac{1}{1 - \alpha} \ln \, \mathrm{Tr }\, \rho_{\textrm{GGE}}^{\alpha}.
\label{eq:sRenyiEntropyDefinition}
\end{equation}
After plugging~\eqref{eq:rGGEDensityMatrix} into~\eqref{eq:sRenyiEntropyDefinition}, 
the Renyi entropies read 
\begin{equation}
S^{(\alpha)}_{\mathrm{GGE}} = \frac{1}{1-\alpha} \left[ \ln  \mathrm{Tr}\, \exp 
  \left(- \alpha \sum_{s,j} \beta^{(j)}_{s}  Q^{(j)}_{s} \right) - \alpha \ln  Z_{\mathrm{GGE}} 
\right].\label{eq:sRenyiEntropyGGE}
\end{equation}
We now review the TBA approach to calculate the GGE R\'enyi entropies. First,  the trace over the eigenstates 
in~\eqref{eq:sRenyiEntropyGGE} in the thermodynamic limit is replaced by a 
functional integral over the TBA densities $\rho_n$ as 
\begin{equation}
\label{th}
\textrm{Tr}\to\int D[\rho]e^{S_{YY}}, 
\end{equation}
where we defined $D[\rho]\equiv\prod_n D\rho_n(\lambda)$. In Eq.~\eqref{th} 
the Yang-Yang entropy takes into account that there is an exponentially 
large (with system size) number of microscopic eigenstates corresponding to 
the same thermodynamic state. The R\'enyi entropies~\eqref{eq:sRenyiEntropyGGE} 
are then given by the functional integral \cite{AlCa17,AlCa17b}
\begin{equation}
S^{(\alpha)}_{\mathrm{GGE}} = \frac{1}{1-\alpha}
\left[ \ln  \int D [\rho] \exp \left(-\alpha 
\mathcal E [\rho]  +S_{\mathrm YY}[\rho]  \right) + 
\alpha  f_{\mathrm{GGE}}\right]. 
\label{eq:sRenyiEntropyTBAFunctionalIntegral}
\end{equation}
Here we introduce the pseudoenergy ${\mathcal E}[\rho]$ as 
\begin{equation}
  \mathcal{E}[\rho] \equiv \sum_{s,j} \beta_{s}^{(j)} Q_{s}^{(j)}[\rho] =
  L\sum_{s,j} \beta_{s}^{(j)} \sum_{n} \int \mathrm{d} \lambda \rho_{n}(\lambda) q_{s,n}^{(j)}(\lambda), 
\label{eq:eEpsilonPseudoEnergy}
\end{equation}
where $q_{s,n}^{(j)}$ are defined in~\eqref{eq:conservedChargeEigenvaluesTBA}.  (In the case of the 
standard Gibbs ensemble the sum includes only the energy $q_{1/2,n}^{(2)}(\lambda)$ coupled with the inverse temperature $\beta_{1/2}^{(2)}$.)
The quantity $f_{\textrm{GGE}}$ is the GGE grand canonical potential defined 
as  
\begin{equation}
  f_{\mathrm{GGE}}=-\ln  Z_{\mathrm{GGE}}. 
  \label{eq:fGGEGrandCanonicalPotential}
\end{equation}

Following the standard TBA treatment~\cite{Taka99}, the functional 
integral in~\eqref{eq:sRenyiEntropyTBAFunctionalIntegral} 
can be evaluated using the saddle-point method~\cite{AlCa17,AlCa17b} 
because both $S_{YY}$ and ${\cal E}$ are extensive. 
Formally, this corresponds to minimising with respect to $\rho_n$ the 
functional ${\mathcal S}^{(\alpha)}_\textrm{GGE}$ defined as 
\begin{equation}
\label{func}
{\mathcal S}^{(\alpha)}_\textrm{GGE}[\rho]\equiv-\alpha 
\mathcal E [\rho]  +S_{\mathrm YY}[\rho]  .
\end{equation}
Notice in~\eqref{func} the explicit dependence on the R\'enyi index $\alpha$. 
For $\alpha=1$, Eq.~\eqref{func} provides the macrostate that describes 
local and quasilocal properties of the steady state~\cite{caux-2013,caux-16} 
and the von Neumann entropy~\cite{alba-2016}. The minimisation procedure gives a set of 
coupled integral equations for 
the macrostate densities $\rho^{(\alpha)}_n$. These are conveniently written in terms 
of  a set of functions $\eta^{(\alpha)}_{n}(\lambda)=\rho^{(\alpha)}_{\mathrm{h},n}
(\lambda)/\rho^{(\alpha)}_{n}(\lambda)$, where $\alpha$ is the index 
of the Renyi entropy. 
Specifically, the saddle point condition on~\eqref{func} yields the equations
\begin{equation}
\ln  \eta^{(\alpha)}_{n} = \alpha g_{n} + \sum_{m=1}^{\infty} A_{nm} 
\star \ln  [1 +1/\eta^{(\alpha)}_{m}],
\label{eq:eRenyiEntropySaddlePointEquationsCoupled}
\end{equation}
where $A_{nm}(\lambda)$ is defined in \eqref{eq:aBGTCoupledKernelANM}, 
and the  TBA driving function $g_{n}(\lambda)$ is defined as 
\begin{equation}
  g_{n}(\lambda) \equiv \sum_{s,j} \beta_{s}^{(j)} q_{s,n}^{(j)}(\lambda).
  \label{eq:gRenyiEntropyCoupledSourcesDefinition}
\end{equation}
Here $q_{s,n}^{(j)}(\lambda)$ are the functions expressing the 
eigenvalues of (quasi)local charges as in \eqref{eq:conservedChargeEigenvaluesTBA}. 
Similarly to the standard TBA~\cite{Taka99}, it is possible to 
partially decouple the system of integral 
equations \eqref{eq:eRenyiEntropySaddlePointEquationsCoupled}, obtaining
\begin{equation}
  \ln  \eta^{(\alpha)}_{n} = \alpha d_{n} + s \star \ln  (1 + \eta^{(\alpha)}_{n-1})(1 + \eta^{(\alpha)}_{n+1})  \qquad (\eta_{0} \equiv 0),
  \label{eq:eRenyiEntropySaddlePointEquationsDecoupled}
\end{equation}
with the source terms $d_n$ being defined as 
\begin{equation}
  d_{n} = g_{n} - s \star (g_{n-1} + g_{n+1})  \qquad (g_{0} \equiv 0).
  \label{eq:RenyiEntropyDecoupledSourcesDefinition}
\end{equation}
This set of equations is easier to solve numerically 
than~\eqref{eq:eRenyiEntropySaddlePointEquationsCoupled} because they contain 
fewer convolutions. Once the solutions $\eta_{n}^{(\alpha)}(\lambda)$ 
are determined, the particle densities $\rho^{(\alpha)}_n$ are 
obtained by using the thermodynamic Bethe 
equations~\eqref{eq:rBetheGaudinTakahashiDecoupled}.
Finally, the GGE Renyi entropy~\eqref{eq:sRenyiEntropyGGE} is obtained 
by evaluating~\eqref{eq:sRenyiEntropyTBAFunctionalIntegral} on the densities 
$\rho^{(\alpha)}_n$ as 
\begin{equation}
S^{(\alpha)}_{\mathrm{GGE}} = \left.\frac{1}{\alpha-1} \Big[\Big( 
\alpha \mathcal E - S_{\mathrm{YY}}\Big)\right|_{\rho_n^{(\alpha)}} + 
\alpha f_{\mathrm{GGE}}\Big|_{\rho_n^{(1)}}\Big],
\label{eq:sRenyiEntropyGGEIntegralExpression}
\end{equation}
where $\mathcal E[\rho]$ and $S_{\mathrm{YY}}[\rho]$ are functionals 
of the string densities defined in \eqref{eq:eEpsilonPseudoEnergy} and 
\eqref{eq:sYangYangEntropy}, and $f_{\mathrm{GGE}}$ is the grand 
canonical potential~\eqref{eq:fGGEGrandCanonicalPotential}. Note that 
$f_{\textrm{GGE}}$ is calculated over the macrostate $\rho_n^{(1)}$, i.e., 
with $\alpha=1$; for all the quenches that can be 
treated with the Quench Action method one has $f_{\textrm{GGE}}=0$, due to the normalisation of
the overlaps. 

Once again, we stress that the thermodynamic macrostate describing the R\'enyi entropies is not the same as that characterising
the local observables, or the von Neumann entropy, and it depends on $\alpha$. 
This has the intriguing consequence that different R\'enyi entropies, in principle, contain information about different 
regions of the spectrum of the post-quench Hamiltonian. 
This difference does not come as a surprise when considering the thermodynamic entropies.  However, the thermodynamic entropies are the same as the entanglement entropies of a subsystem that is large in itself but a vanishing fraction of the whole system. Therefore the difference is very puzzling because entanglement entropies are all calculated from the same quantum mechanical wavefunction.

A technical remark is now in order. 
Although the procedure to extract the R\'enyi entropies that we outlined so far 
is legitimate, a crucial ingredient is the set of infinitely many Lagrange 
multipliers $\beta_{s}^{(j)}$ entering in \eqref{eq:gRenyiEntropyCoupledSourcesDefinition} 
and in the driving functions $g_n$ (cf.~\eqref{eq:gRenyiEntropyCoupledSourcesDefinition}). 
In principle, they are fixed by requiring that the GGE averages of the local and 
quasilocal conserved quantities $Q_s^{(j)}$ equal their expectation values over the 
initial state. 
However, this is a formidable task that cannot be carried out in practice.  
As of now, it is possible to overcome this difficulty in two ways. One is to use 
the  Quench Action method~\cite{caux-2013,caux-16}, as discussed below. 
The other way, based on the analytical solution of the GGE saddle point 
equations, will be described in Section~\ref{sec:sourceTermExtraction}.
The Quench Action 
gives access to the thermodynamic macrostate describing local and 
quasilocal observables (and the thermodynamic entropy) in the stationary state at 
long times. Crucially, the Quench Action allows one to extract the driving 
functions $g_n$ without relying on the knowledge of the $\beta_s^{(j)}$. The key 
ingredients are the overlaps between the initial state and all the eigenstates of 
the post-quench Hamiltonian, although a subset of the thermodynamically relevant 
overlaps may be  sufficient (see~\cite{wdbf-14,PMWK14,ac-16qa}). 
Crucially, for a large class of initial states the overlaps can be calculated 
analytically \cite{fcc-09,Pozsgay18,grd-10,pozsgay-14,dwbc-14,bdwc-14,cd-12,pc-14,fz-16, dkm-16,hst-17,msca-16,pe-16,BeSE14,BePC16,BeTC17}. 
Overlap calculations are possible also for systems in the continuum, such as the Lieb-Liniger gas. 
For example, in Refs. \cite{dwbc-14,bdwc-14,cd-12}, 
the overlaps between the Bose condensate (BEC) state and the eigenstates of the 
Lieb-Liniger model with both repulsive and attractive interactions, have been calculated
and used in \cite{dwbc-14,pce-16} to study their dynamics. 
The information about the driving functions was crucial in Ref.~\cite{AlCa17b} 
to obtain the steady-state value of the R\'enyi entropies after the quench 
from the N\'eel state. 
Interestingly, all the initial states for which it has been 
possible to calculate the overlaps in interacting models are reflection symmetric. 
The defining property of reflection-symmetric states is that they have nonzero overlap only 
with parity-invariant eigenstates, which correspond to solutions of the 
Bethe equations that contain only pairs of rapidities with opposite sign, i.e., 
such that $\{\lambda_j\}_{j=1}^N=\{-\lambda_j\}_{j=1}^N$. 
Indeed, the importance of parity-invariance for the solvability of quantum quenches 
has been explored in Ref. \cite{ppv-17} for integrable lattice models 
and in Ref. \cite{delfino-14}  for integrable field theories. 
However, some non reflection symmetric initial states 
for which the Quench Action method can be applied have been found in the 
Hubbard chain in the infinite repulsion limit in Ref.~\cite{BeTC17}.


\subsection{A simplified expression for the R\'enyi entropies}
\label{sec-simp}

The GGE R\'enyi entropy as expressed in \eqref{eq:sRenyiEntropyGGEIntegralExpression} 
are functionals of an infinite set of densities $\rho_{n}(\lambda)$. 
In this section we show that it is possible to simplify~\eqref{eq:sRenyiEntropyGGEIntegralExpression} 
writing the R\'enyi entropies only in terms of $\eta_1^{(\alpha)}$. 
A formula similar to the one we are going to derive is known for the Gibbs (thermal) free energy
since many years~\cite{Taka99}. 

The first step in this derivation is to rewrite~\eqref{eq:sRenyiEntropyGGEIntegralExpression} as 
\begin{align}
S^{(\alpha)}_{\mathrm{GGE}} &= \frac{L}{\alpha-1}\sum_{n=1}^{\infty} \int_{-\pi/2}^{\pi/2}\mathrm d \lambda\bigg[ 
	\alpha \rho^{(\alpha)}_{n}(\lambda) g_{n}(\lambda)
	- \rho_{n}^{(\alpha)}(\lambda) \ln (1 + \eta^{(\alpha)}_{n}(\lambda)) 
- \rho_{\mathrm{h},n}^{(\alpha)}(\lambda) \ln (1 + 1/\eta_{n}^{(\alpha)}(\lambda))\bigg]+
   \frac{\alpha L}{\alpha-1} f_{\mathrm{GGE}}.
  \label{eq:sRenyiEntropyGGEIntegralExpressionFull}
\end{align}
The function $\alpha g_{n}(\lambda)$ can be obtained from Eq.~\eqref{eq:eRenyiEntropySaddlePointEquationsCoupled}, 
whereas $\rho_{\mathrm{h,}n}^{(\alpha)}(\lambda)$ can be obtained from the BGT 
equations in~\eqref{eq:rBetheGaudinTakahashiCoupled}. 
After some algebra this yields 
\begin{equation}
 S^{(\alpha)}_{\mathrm{GGE}} = \frac{L}{1-\alpha} \left[ \sum_{n=1}^{\infty} \int_{-\pi/2}^{\pi/2}\! \mathrm{d} \lambda\, a_{n}(\lambda) \ln (1 + 1/\eta_{n}^{(\alpha)}(\lambda)) - \alpha f_{\mathrm{GGE}} \right]. 
  \label{eq:sRenyiEntropyStrangeSum}
\end{equation}
The infinite sum in~\eqref{eq:sRenyiEntropyStrangeSum} 
can be further simplified by considering the first of the 
saddle point equations in~\eqref{eq:eRenyiEntropySaddlePointEquationsCoupled}
\begin{equation}
  \ln (1 + \eta^{(\alpha)}_{1}(\lambda)) = g_{1}(\lambda) + \sum_{m=1}^{\infty} \int_{-\pi/2}^{\pi/2} \! \mathrm d \mu \, \left[a_{m-1}(\lambda-\mu) + a_{m+1}(\lambda-\mu)\right]\, \ln (1 + 1/\eta^{(\alpha)}_{m}(\mu)). 
  \label{eq:eFirstSaddlePointEquation}
\end{equation}
One then has to multiply~\eqref{eq:eFirstSaddlePointEquation} 
by $s(\lambda)$ (cf.~\eqref{eq:sFunction}) and integrate over 
$\lambda$. Finally, after some manipulations (identical to those appearing in Ref.~\cite{wdbf-14}), one obtains   
\begin{equation}
  \int_{-\pi/2}^{\pi/2}\! \mathrm{d} \lambda \, s(\lambda) \left[ \ln  (1 + \eta^{(\alpha)}_{1}(\lambda)) - g_{1}(\lambda) \right] = 
  \sum_{n=1}^{\infty} \int_{-\pi/2}^{\pi/2}\! \mathrm{d} \lambda \, a_{n}(\lambda) \ln (1 + 1/\eta_{n}^{(\alpha)}(\lambda)).
  \label{eq:eFirstSaddlePointEquationConvolved} 
\end{equation}
The right-hand side of~\eqref{eq:eFirstSaddlePointEquationConvolved} 
is precisely the first term in the square brackets in~\eqref{eq:sRenyiEntropyStrangeSum}.
Plugging~\eqref{eq:eFirstSaddlePointEquationConvolved} 
into~\eqref{eq:sRenyiEntropyStrangeSum}, one obtains the simplified formula 
for the Rényi entropy as 
\begin{equation}
  S^{(\alpha)}_{\mathrm{GGE}} = \frac{L}{\alpha-1} \left\{ \int_{-\pi/2}^{\pi/2} \mathrm{d}\lambda\, s(\lambda) \left[ \alpha g_{1}(\lambda) - \ln (1+\eta^{(\alpha)}_{1}(\lambda)) \right] + \alpha f_{\mathrm{GGE}} \right\},
  \label{eq:sRenyiEntropyTheta1Expression}
\end{equation}
which is our final result depending only on $\eta^{(\alpha)}_{1}(\lambda)$.

An important remark is that while~\eqref{eq:sRenyiEntropyTheta1Expression} 
depends only on one rapidity density, it is still necessary to solve the full 
set of TBA equations~\eqref{eq:eRenyiEntropySaddlePointEquationsCoupled} 
in order to determine $\eta^{(\alpha)}_{1}$, because all the 
$\eta_n^{(\alpha)}$ are coupled. 
However,  Eq.~\eqref{eq:sRenyiEntropyTheta1Expression} has at least two 
advantages. First, it is more convenient than~\eqref{eq:sRenyiEntropyGGEIntegralExpression} 
from the numerical point of view, because it contains less integrals to be 
evaluated. Second, Eq.~\eqref{eq:sRenyiEntropyTheta1Expression} is more 
convenient for analytical manipulations. 

\subsection{R\'enyi entropies after quenching from the dimer state}
\label{sec:dimer}

In this section we employ the TBA approach described  above
to calculate the R\'enyi entropies after the quench from the dimer state, 
generalising the results of Ref.~\cite{AlCa17b}  for the quench from the N\'eel state. 
For the dimer state, the overlaps are analytically known \cite{PMWK14,pozsgay-14} and hence the Quench Action provides  
the driving functions  $g_n$ as \cite{PMWK14}
\begin{align}
  \begin{split}
    g_{1}(\lambda) &= -\ln  \left(\frac{\sinh^{4}(\lambda) \cot^{2}(\lambda)}{\sin(2 \lambda + i\eta)\, \sin(2 \lambda- i\eta)}\right), \\
    g_{n}(\lambda) &= \sum_{k=1}^{n} g_{1}(\lambda+i \eta (n+1-2k)/2 ), \qquad (n \ge 2),
    \label{eq:gSourceFunction}
  \end{split} \\
   d_{n}(\lambda) &= -\ln \! \left( \frac{ \vartheta_{4}(\lambda)}{ \vartheta_{1}(\lambda)} \right)^2 + (-1)^{n}\ln \! \left( \frac{ \vartheta_{2}(\lambda)}{ \vartheta_{3}(\lambda)} \right)^{2},
  \label{eq:gDimerSourceFunctionD}
\end{align}
where $\vartheta_{\ell}(x)$ are the Jacobi $\vartheta$-functions with nome $e^{-2 \eta}$.

The strategy to calculate the R\'enyi entropies is to use the driving function $g_n$  
in the TBA equations for $\eta_n^{(\alpha)}$ (cf.~\eqref{eq:eRenyiEntropySaddlePointEquationsCoupled}). 
After solving for $\eta^{(\alpha)}_n$, the GGE R\'enyi entropies are obtained from~\eqref{eq:sRenyiEntropyTheta1Expression}. 
Obviously, the term $f_{\textrm{GGE}}$ has 
to be evaluated using the density $\rho_n^{(1)}$ (cf. ~\eqref{eq:sRenyiEntropyGGEIntegralExpression}), but normalisation 
provides $f_{\textrm{GGE}}=0$. 

\begin{figure}[t]
\centering
\includegraphics{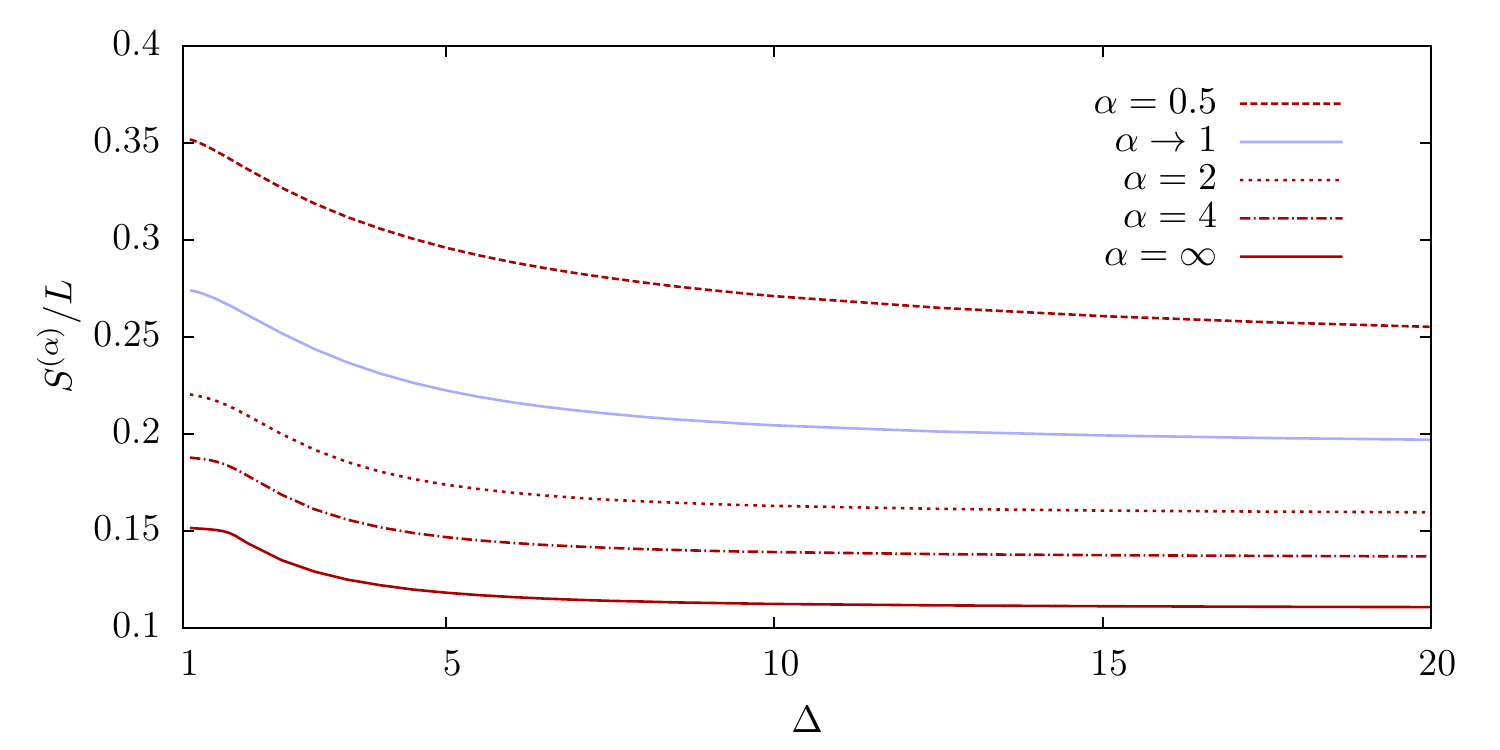}
\caption{R\'enyi  entropies of the GGE after the  quench from the dimer state in the XXZ chain.
The entropy densities $S^{(\alpha)}/L$ are plotted as a function of the chain anisotropy $\Delta$. 
The different lines denote results for different R\'enyi index $\alpha$. 
In the limit $\Delta\to\infty$ all the R\'enyi entropies remain finite. 
}
\label{fig:dimerRenyiEntropies}
\end{figure}

The numerical results for the R\'enyi entropies obtained with this procedure 
are shown in Figure~\ref{fig:dimerRenyiEntropies}. The Figure shows the entropy 
densities $S^{(\alpha)}/L$ plotted versus the chain anisotropy $\Delta$. Different lines 
correspond to different values of $\alpha$. As expected, one has that for 
any $\Delta$, $S^{(\alpha)}<S^{(\alpha')}$ if $\alpha>\alpha'$. 
For completeness we report the result for $\alpha\rightarrow1$. 
An interesting observation is that the R\'enyi entropies do not vanish in the Ising limit for 
$\Delta\rightarrow\infty$. This is in contrast  with what happens for the 
quench from the N\'eel state, for which the steady-state entropies at $\Delta\to\infty$ vanish. 
The reason is that the N\'eel state becomes the ground state of the XXZ chain in that limit, and there is no 
dynamics after the quench. 
In some limiting cases it is possible to derive closed analytic formulas for the post-quench stationary 
R\'enyi entropy: in the following, we will provide analytical results 
in the Ising limit $\Delta\rightarrow\infty$ and for the min entropy, i.e., the limit $\alpha\rightarrow\infty$.  

\subsection{Expansion of Rényi entropies in the Ising limit}
\label{sub:HighDeltaExpansion}

In this section we perform an expansion of the steady-state R\'enyi 
entropies in the large $\Delta$ limit, by closely following the procedure introduced in~\cite{wdbf-14}. 
A similar expansion for the R\'enyi entropies 
after the quench from the N\'eel state has been carried out in~\cite{AlCa17}. 
In that case, the $\Delta\rightarrow\infty$ limit is very special, 
since the N\'eel state becomes the ground state of the model, and there is no dynamics. 
The quench  from the dimer state is more generic, because the dimer state is never an eigenstate of 
the chain, and consequently the post-quench dynamics is nontrivial, implying that the 
stationary value of the R\'enyi entropy is nonzero also for $\Delta\to\infty$.
We anticipate that in the Ising limit the R\'enyi 
entropies have the same form as for free-fermion models~\cite{AlCa17}, but 
deviations from the free-fermion result appear already at 
the first non trivial order in $1/\Delta$. 

In the following we obtain the steady-state R\'enyi entropies 
as a power series in the variable $z\equiv e^{-\eta}$ with 
$\eta=\textrm{arccosh}(\Delta)$. 
Following~\cite{AlCa17b}, we use the ansatz for $\eta^{(\alpha)}_n$ 
\begin{align}
\eta_{n}^{(\alpha)}(\lambda) = z^{\beta_{n}(\alpha)} \eta^{(\alpha)}_{n,0}(\lambda) \exp \left(\Phi^{(\alpha)}_{n}(z,\lambda) \right). 
  \label{eq:eLargeDeltaAnsatz}
\end{align}
Here the exponents $\beta_n(\alpha)$, the functions 
$\Phi^{(\alpha)}_n(\lambda)$, and $\eta^{(\alpha)}_{n,0}$ 
have to be determined by plugging the 
ansatz~\eqref{eq:eLargeDeltaAnsatz} into the TBA 
equations~\eqref{eq:eRenyiEntropySaddlePointEquationsDecoupled}. 
We also need the expansion of the driving functions $d_n$ around $z=0$:
\begin{align}
  d_{n}=
  \begin{cases}
    \ln (4\sin^{2}(2\lambda))\, z^{2} + 4 \cos(4\lambda)\, z^{4} + O(z^{6})& n \text{ even,}\\
    \ln (\tan^{2}(\lambda))\, + 8\cos(2\lambda)\, z^{2} - 8\cos(2\lambda)\, z^{4} + O(z^{6}) & n \text{ odd}. 
  \end{cases}
  \label{eq:dExpansion}
\end{align}
The expansion of the kernel $s(\lambda)$ appearing in~\eqref{eq:eRenyiEntropySaddlePointEquationsDecoupled} is
\begin{equation}
s(\lambda) = \frac{1}{2\pi} + \frac{2}{\pi} \cos(2\lambda) z + \frac{2}{\pi} \cos (4\lambda) z^{2} + O(z^3). 
\label{eq:sExpansion}
\end{equation}
After plugging the ansatz~\eqref{eq:eLargeDeltaAnsatz} 
into~\eqref{eq:eRenyiEntropySaddlePointEquationsDecoupled}, and considering 
the leading order in powers of $z$, one can fix the exponents $\beta_n$. 
By treating separately the cases of even and odd $n$ in~\eqref{eq:dExpansion}, one finds
\begin{equation}
  \beta_{n} =
  \begin{cases}
    2\alpha & n \text{ even,} \\
    0 & n \text{ odd}.
  \end{cases}
  \label{eq:betaExponents}
\end{equation}
This choice is not unique, but it is the only one that is 
consistent with the BGT equations 
\eqref{eq:rBetheGaudinTakahashiDecoupled}, see~\cite{wdbf-14}.
The leading order in $z$ of~\eqref{eq:eRenyiEntropySaddlePointEquationsDecoupled} 
 fixes the functions $\eta_{n,0}(\lambda)$ as
\begin{equation}
	\eta_{n,0}^{(\alpha)}(\lambda) = 
  \begin{cases}
    4^{\alpha}\,e^{c(\alpha)}\, |\sin(\lambda)|^{2\alpha} & n \text{ even,} \\
    |\tan(\lambda)|^{2\alpha} & n \text{ odd,}
  \end{cases}
  \label{eq:etaExpansionLeadingCoefficients}
\end{equation}
where the constant $c(\alpha)$ is given as 
\begin{equation}
  c(\alpha) = \frac{1}{\pi} \int_{-\pi/2}^{\pi/2}\! \mathrm d \mu \ln  (1 + |\tan(\mu)|^{2\alpha}).
  \label{eq:cExpansionLeadingCoefficientConstant}
\end{equation}
In the limit $\alpha \rightarrow 1$, one has $c(1)=\ln 4$. 
Interestingly, Eq.~\eqref{eq:etaExpansionLeadingCoefficients} shows
that for $n$ odd, $\eta_n^{(\alpha)}$ is a regular function for 
any value of $\lambda$, whereas for even $n$ it diverges for $\lambda=\pm\pi/2$. 
This is a striking difference compared to the quench from the N\'eel state, for which 
$\eta_n^{(\alpha)}$ diverges in the limit $\lambda\to 0$ for even 
$n$ (see Ref.~\cite{AlCa17b}), whereas it is regular for odd $n$. 

Also, at the leading order in $z$, one has that $\Phi_n^{(\alpha)}=0$.  
By combining the results in~\eqref{eq:etaExpansionLeadingCoefficients} 
and~\eqref{eq:betaExponents} with the TBA equations~\eqref{eq:rBetheGaudinTakahashiDecoupled}, 
the leading order of the rapidity densities  $\rho^{(\alpha)}_{\textrm{t},n}$ are
\begin{align}
	\label{eq:r1}
	\rho_{\textrm{t},1}^{(\alpha)} &= \frac{1}{2\pi}(1 + 4\cos(2\lambda)\, z) + O(z^{\min (2,2\alpha)}), \\
	\label{eq:r2}
	\rho_{\textrm{t},2}^{(\alpha)} &= \frac{1}{8\pi} + O(z^{\min (1,2\alpha)}), \\
	\rho_{\textrm{t},n}^{(\alpha)} &= O(z^{2\alpha}) \qquad (n\ge 2).  
  \label{eq:r3}
\end{align}

Notice that for any $\alpha$, only $\rho_n^{(\alpha)}$ and $\rho_{\textrm{t},n}^{(\alpha)}$ with $n\le2$ remain finite in the limit $z\to0$, 
whereas all densities with $n>2$ vanish. This is different from the N\'eel state, 
for which only the densities with $n=1$ are finite~\cite{AlCa17b}. 
Physically, this is expected because in the dimer state only components 
with at most two aligned spins can be present. 
Furthermore, 
the leading order of $\rho_{\textrm{t},n}^{(\alpha)}$ 
in~\eqref{eq:r1}-\eqref{eq:r3} does not depend on the R\'enyi index $\alpha$. 
Finally, in the limit $z\to0$, the densities 
become constant, similar to free-fermion models. This suggests that in the 
limit $z\to0$ the form of the R\'enyi entropies may be similar to that of
free models, as we are going to show in the following. 

It is now straightforward to derive the leading behaviour of the 
R\'enyi entropies for any $\alpha$ in the limit $z\to0$. First, we obtain 
the expansion of the driving functions $g_n$~\eqref{eq:gSourceFunction} as 
\begin{align}
  g_{1}(\lambda) &= \ln  (4 \tan^{2}(\lambda)) + 4z + 4 \sin^{2}(2\lambda) z^2 + O(z^{3}), 
  \label{eq:1}\\
  g_{2}(\lambda) &= \ln  (64 \sin^{2}(2 \lambda)) + 16\sin^{2}(\lambda)\,z + 4 z^{2} + O(z^{3}).
  \label{eq:2}
\end{align}
The contributions of $g_n$ for $n>2$ are subleading for $z\to0$ and may be neglected. 
Using the expansions~\eqref{eq:1}~\eqref{eq:2}, the leading order 
of the densities in~\eqref{eq:r1}-\eqref{eq:r3}, and~\eqref{eq:etaExpansionLeadingCoefficients}, 
one obtains the $z \rightarrow 0$ limit of ${\cal E}$ (cf.~\eqref{eq:eEpsilonPseudoEnergy}) 
as 
\begin{equation}
\begin{split}
\mathcal{E} &= \frac{L}{4\pi} \Big[\int_{0}^{\pi/2}\! \mathrm d\lambda\, \frac{\ln(4|\tan(\lambda)|^{2}) }
{1 + |\tan(\lambda)|^{2\alpha}} + \frac{1}{4} \int_{0}^{\pi/2}\!\mathrm d\lambda\, \ln(64|\sin(2 \lambda)|^{2}) 
\Big]= \frac{L}{8}\,\ln(2) +\frac{L}{4\pi} \int_{0}^{\pi/2}\! \mathrm d\lambda\, \frac{\ln(4|\tan(\lambda)|^{2})}{1 + |\tan(\lambda)|^{2\alpha}}.
\end{split}
  \label{eq:ePseudoEnergyZ0Limit}
\end{equation} 
The two terms in the right-hand-side of~\eqref{eq:ePseudoEnergyZ0Limit} 
are the contributions of the densities with $n=1$ and $n=2$. 
Similarly, the $z\rightarrow 0$ limit of the Yang-Yang 
entropy~\eqref{eq:sYangYangEntropy} is obtained as 
\begin{equation}
  S_{\mathrm{YY}} = \frac{L}{2 \pi}\, \int_{-\pi/2}^{\pi/2}\!\mathrm d\lambda \left( \frac{1}{1+|\tan(\lambda)|^{2 \alpha}}\ln (1 + |\tan(\lambda)|^{2 \alpha}) + \frac{1}{1+|\cot(\lambda)|^{2 \alpha}} \ln (1 + |\cot(\lambda)|^{2 \alpha}) \right).
  \label{eq:sYangYangEntropyZ0Limit}
\end{equation}
Plugging Eq.~\eqref{eq:ePseudoEnergyZ0Limit}-\eqref{eq:sYangYangEntropyZ0Limit} into 
the definition of the R\'enyi entropies \eqref{eq:sRenyiEntropyDefinition}, 
one obtains at the leading order in $z$ 
\begin{equation}
	S^{(\alpha)}_{\textrm{GGE}} = \frac{L}{1-\alpha} \int_{-\pi/2}^{\pi/2}\!\frac{\mathrm d
    \lambda}{2 \pi}  \ln \left[\left( \frac{1}{1+\tan^{2}(\lambda)} 
    \right)^{\alpha}+\left(1 -  \frac{1}{1+\tan^{2}(\lambda)} \right)^{\alpha}\right].
  \label{eq:sRenyiEntropyZ0Limit}
\end{equation}
Eq.~\eqref{eq:sRenyiEntropyZ0Limit} shows that for any $\alpha$ the R\'enyi entropies 
are not vanishing in the limit $z\to0$. 

\begin{figure}[t]
  \centering
  \includegraphics{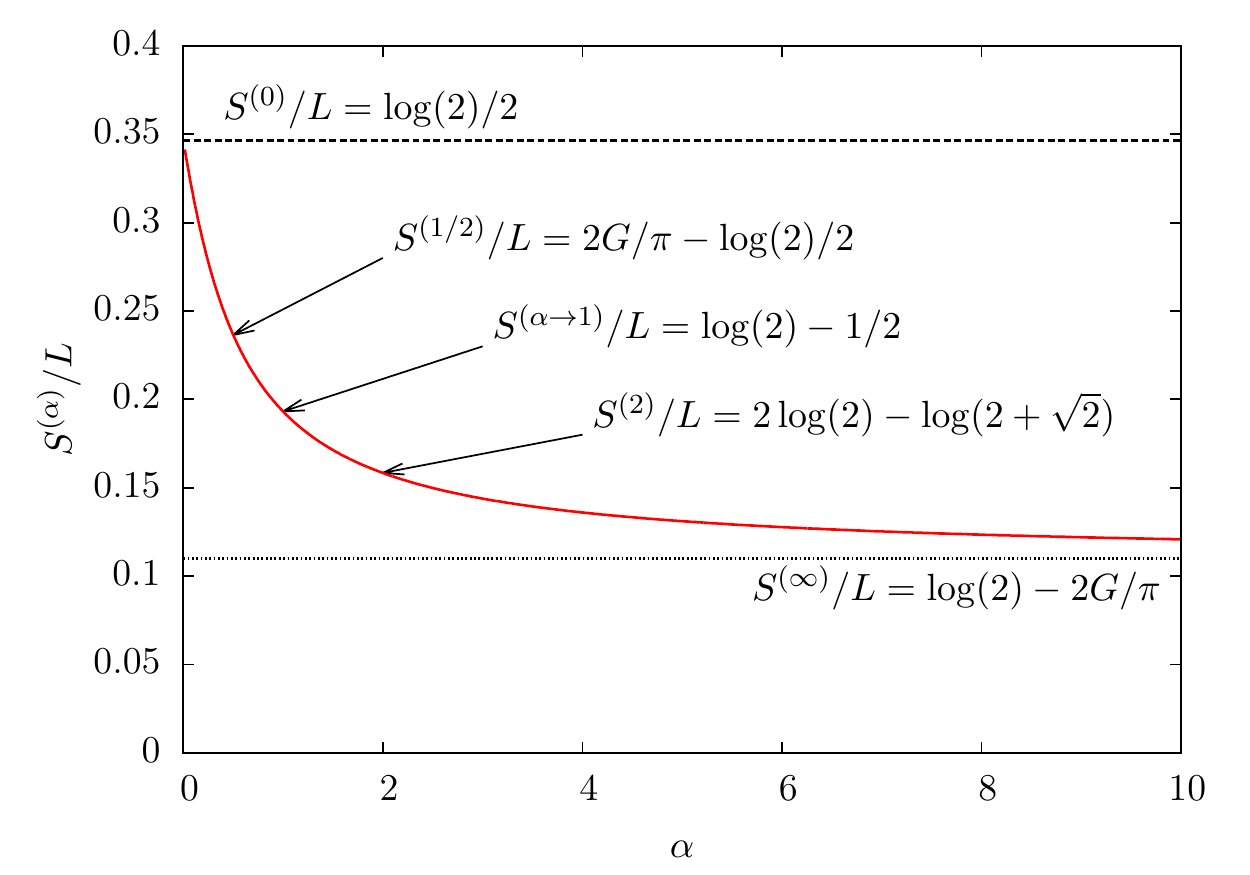}
  \caption{ R\'enyi entropy densities $S^{\alpha}/L$ after the quench 
  from the dimer state in the Ising limit $\Delta\to\infty$ of the XXZ 
  chain. On the $x$-axis $\alpha$ is the R\'enyi index. The results are 
  obtained using~\eqref{eq:sRenyiEntropyZ0Limit}. The dashed and the 
  dotted lines show the max entropy $S^{(0)}$ and the 
  min entropy $S^{(\infty)}$, respectively. 
  Here $G$ is the Catalan constant.
 }
  \label{fig:sRenyiEntropiesIsingLimit}
\end{figure}

The R\'enyi entropy obtained from~\eqref{eq:sRenyiEntropyZ0Limit} are plotted in Fig.~\ref{fig:sRenyiEntropiesIsingLimit} 
as a function of the R\'enyi index $\alpha$. 
Like for finite $\Delta$, the R\'enyi entropies are monotonically decreasing functions of $\alpha$. 
For some values of $\alpha$ the integrals in \eqref{eq:sRenyiEntropyZ0Limit} can be computed analytically. 
For instance, for the max entropy, i.e., in the limit $\alpha\to0$, one obtains $S^{(0)}/L = \ln(2)/2$:
since the max entropy is twice the logarithm of the total number of eigenstates that have nonzero 
overlap with the initial state \cite{AlCa17}, we have that  that this number is $\propto e^{L/2}$ (which is  
the same result for the quench from the N\'eel state~\cite{AlCa17b}). 
For the min entropy we have $S^{(\infty)}/L = \ln(2) - 2G/\pi$, where $G$ is the Catalan constant. 
Some other analytical results  are reported in the Figure.

It is relatively easy to obtain the higher-order corrections in powers of $z$ for any {\it fixed} $\alpha$. 
Instead, it is rather cumbersome to carry out the expansion for general real $\alpha$. 
Therefore, here we only show the explicit calculation for the case $\alpha = 2$. 
Specifically, we determine $S^{(2)}$ up to ${\mathcal O}(z^2)$. For convenience, instead of $\eta_n^{(\alpha)}$ we 
consider the filling functions $\vartheta_n\equiv 1/(1+\eta_n)$, where we 
suppressed the dependence on $\alpha$, because we consider $\alpha=2$. 
The derivation of the higher-order expansion for the filling functions is the same as in Ref.~\cite{AlCa17b} and 
we will omit it. The idea is that one has to plug the ansatz~\eqref{eq:eLargeDeltaAnsatz} 
into the equations for $\eta_n$ (cf.~\eqref{eq:eRenyiEntropySaddlePointEquationsDecoupled}) 
solving the system order by order in powers of $z$. Similar to the 
N\'eel quench, we observe that the system~\eqref{eq:eRenyiEntropySaddlePointEquationsDecoupled} 
contains an infinite set of equations (one for each string type). However, 
to obtain the filling functions up to terms ${\mathcal O}(z^\omega)$, only 
the first $m(\omega)\sim \omega$ equations matter because the leading order of higher strings is given by higher 
orders in powers of $z$. The expansion of the filling functions $\vartheta_n$ around $z=0$ reads 
\begin{align}
  \vartheta_{1}(\lambda) &= \frac{1}{1 + \tan(\lambda)^{4}(\lambda)} - \frac{16 \cos(2\lambda)\tan^{4}(\lambda)}{(1 + \tan^{4}(\lambda))}z^{2} + O(z^{3}),
  \label{eq:theta1SmallZExpansion} \\
  \vartheta_{2}(\lambda) &= 1 + O(z^{4}),
  \label{eq:theta2SmallZExpansion}\\
  \vartheta_{n}(\lambda) &= O(z^{4}), \qquad (n \ge 3).
  \label{eq:thetaLargerNSmallZExpansion}
\end{align}
A similar procedure for the TBA equations for the particle 
densities (cf.~\eqref{eq:rBetheGaudinTakahashiDecoupled}) gives 
\begin{align}
  \rho_{\mathrm t,1} &= \frac{1}{2 \pi} + \frac{2}{ \pi}\cos(2\lambda) z + \frac{2}{ \pi} \cos(4 \lambda) z^{2} + O(z^{3}),
  \label{eq:rho1tSmallZExpansion}\\
  \rho_{\mathrm t,2} &= \frac{1}{8 \pi} + \frac{2 - \sqrt{2}}{ \pi} \cos^{2}(\lambda) z + \frac{1}{ \pi} \cos (2 \lambda) z ^{2},
  \label{eq:rho2tSmallZExpansion}\\
  \rho_{\textrm{t},n} &= O(z^{4}), \qquad (n \ge 3).
  \label{eq:rhoLargerNtSmallZExpansion}
\end{align}
The next-to-leading order in powers of $z$ of the R\'enyi entropies  
can be computed by plugging~\eqref{eq:sExpansion}\eqref{eq:1}\eqref{eq:2} 
and \eqref{eq:theta1SmallZExpansion}-\eqref{eq:rhoLargerNtSmallZExpansion} 
into~\eqref{eq:eEpsilonPseudoEnergy} and \eqref{eq:sYangYangEntropy}. 
Given that  the first-order in $z$ cancels in both ${\mathcal E}$ and the Yang-Yang entropy and also that  
the  ${\mathcal O}(z^2)$ contribution to $S_{YY}$ vanishes, the first nonzero contribution is 
${\mathcal O}(z^2)$ in ${\mathcal E}$, i.e. 
\begin{equation}
	\mathcal{E}= \frac{1}{4}\ln2 - \frac{\sqrt{2} \pi}{32} + \frac{z^{2}}{2} 
	+ O(z^{3}).
  \label{eq:ePseudoEnergySmallZCorrection}
\end{equation}
Thus, putting the various pieces together, the second R\'enyi entropy $S^{(2)}$ is given as 
\begin{equation}
	S^{(2)} = 2\ln2 - \ln(2+\sqrt{2}) + 2z^{2} + O(z^{3}).
  \label{eq:sRenyiEntropySmallZCorrection}
\end{equation}
Eq.~\eqref{eq:sRenyiEntropySmallZCorrection} implies that the 
asymptotic value of $S^{(2)}$ for $\Delta\to\infty$ is approached 
as $1/\Delta$.

\subsection{A tempting but wrong conjecture}
\label{conje}

It is tempting to investigate the structure of Eq.~\eqref{eq:sRenyiEntropyZ0Limit} which  has the same structure as the 
R\'enyi entropy of free-fermion models, written usually as 
\begin{equation}
S^{(\alpha)}_{\textrm{GGE}} = \frac{L}{1-\alpha} \int_{-\pi/2}^{\pi/2}\!
\frac{\mathrm d k}{2 \pi}  \ln \left[\vartheta({k})^{\alpha}+
\left( 1 - \vartheta({k}) \right)^{\alpha}\right],
  \label{eq:RenyiEntropyLimit}
\end{equation}
where $\vartheta(k)$ are now the free-fermion occupation numbers identifying 
the macrostate. Eq.~\eqref{eq:RenyiEntropyLimit} 
is the same as~\eqref{eq:sRenyiEntropyZ0Limit} after defining  
$\vartheta(k)=1/(1+\tan^2 (k))$. Also,  the factor $1/(2\pi)$ 
in~\eqref{eq:sRenyiEntropyZ0Limit} is the fermionic total density of states 
$\rho_{t}=1/(2\pi)$. A natural question is whether the 
free-fermion formula~\eqref{eq:RenyiEntropyLimit} holds true 
beyond the leading order in $z$. 
For instance, it is interesting to check 
whether~\eqref{eq:sRenyiEntropySmallZCorrection} can be written in 
the free-fermion form~\eqref{eq:RenyiEntropyLimit}. 
However, as the XXZ chain is interacting, Eq.~\eqref{eq:RenyiEntropyLimit} 
requires some generalisation. First, as there are different 
families of strings it is natural to sum over the 
string content of the macrostate. Moreover, in contrast with free 
fermions, for Bethe ansatz solvable models the total density of states 
$\rho_t$ is not constant, but it depends on the string type. 
The most natural generalisation of the free-fermion formula~\eqref{eq:RenyiEntropyLimit} would be 
\begin{equation}
\label{ff-last}
S^{(\alpha)}_{\textrm{GGE}}\stackrel{?}{=}\frac{L}{1-\alpha}\sum_n\int_{-\pi/2}^{\pi/2}
d\lambda\rho_{n,t}\ln[\vartheta_{n}^\alpha+(1-\vartheta_n)^\alpha]. 
\end{equation}
Eq.~\eqref{ff-last} is the same as the free-fermion 
formula~\eqref{eq:RenyiEntropyLimit} except for the overall term $\rho_{\textrm{t},n}$ 
in the integrand, which takes into account that for interacting models 
the density of states is not constant. In Eq.~\eqref{ff-last}, the 
filling functions are given in~\eqref{eq:theta1SmallZExpansion}-\eqref{eq:thetaLargerNSmallZExpansion}. 

Unfortunately, Eq.~\eqref{ff-last} does not give the correct value for the steady-state R\'enyi entropies. 
A very simple counterexample is provided by the standard Gibbs ensemble at infinite temperature. 
For the $XXZ$ chain the macrostate describing this ensemble can be derived using  the standard TBA approach (see~\cite{Taka99}). 
In particular, for the XXX chain the exact infinite temperature R\'enyi entropies can be worked out analytically: 
they become equal and their  density is $S^{(\alpha)}=L \ln2$ for any $\alpha$. 
However, by using the analytical expression~\cite{Taka99} for the infinite-temperature 
filling functions $\vartheta_n$ for the XXX chain in Eq.~\eqref{ff-last}, one can verify that $S^{(2)}/L\ne\ln2$. 

Still, since the free-fermion formula~\eqref{ff-last} holds true exactly 
at $\Delta\to\infty$ (cf.~\eqref{eq:sRenyiEntropySmallZCorrection}), 
it is natural to wonder at which order in $1/\Delta$ (equivalently in $z$) it breaks down. 
By using  Eq.~\eqref{eq:rho1tSmallZExpansion}-\eqref{eq:rhoLargerNtSmallZExpansion} 
and~\eqref{eq:theta1SmallZExpansion}-\eqref{eq:thetaLargerNSmallZExpansion} 
in~\eqref{ff-last}, one can check that $\rho_{\textrm{t},2}$ (cf. Eq.~\eqref{eq:rho2tSmallZExpansion}) 
gives rise to an ${\mathcal O}(z)$ term in the entropy, which is absent  
in~\eqref{eq:sRenyiEntropySmallZCorrection}. This shows that~\eqref{ff-last} 
breaks down already at the first nontrivial order beyond the Ising limit. 

\subsection{The min entropy}
\label{sec-min-dimer}

In this section we discuss the R\'enyi entropy in the limit $\alpha\rightarrow\infty$ also known as 
min entropy,  for which analytical results are obtainable. 
The analysis of the min entropy after the dimer 
quench is similar to that for the N\'eel quench~\cite{AlCa17b}. 
In the following we remove the dependence on $\alpha$ in the 
saddle point densities to simplify the notation. 
After defining the functions $\gamma_{n}=\ln (\eta_{n})/\alpha$, 
the $\alpha\rightarrow\infty$ limit of the saddle-point 
equations~\eqref{eq:eRenyiEntropySaddlePointEquationsDecoupled} yields 
\begin{equation}
  \gamma_{n} = d_{n} + s \star (\gamma_{n-1}^{+}+\gamma_{n+1}^{+}),
  \label{eq:gSaddlePointEquationsAlphaInfinity}
\end{equation}
where $\gamma_{n}^{+}=(\gamma_{n}+|\gamma_{n}|)/2$. 
Some insights on the structure of the solutions 
of~\eqref{eq:gSaddlePointEquationsAlphaInfinity} can be obtained 
by looking at the limit $\Delta\to\infty$. Precisely, 
from Eq.~\eqref{eq:etaExpansionLeadingCoefficients} one has 
that $\eta_n\to 0$ for even $n$ in the limit $\alpha
\to\infty$. On the other hand, one has that $\eta_n$ 
diverges for odd $n$ for $\lambda\in[-\pi/4,\pi/4]$, which 
implies $\ln\eta_n=\alpha d_n$ for $n$ odd in that 
interval. Thus we have that at large $\Delta$, $\gamma_{2n}
(\lambda)<0$ and $\gamma_{2n+1}(\lambda)=d_{2n+1}(\lambda)$. 
As a consequence, the filling functions $\vartheta_n$ become 
\begin{align}
\label{theta1}
  \vartheta_{2n}(\lambda)&= \lim_{\alpha\rightarrow\infty} \frac{1}{1+e^{\alpha\gamma_{2n}(\lambda)}} = 1, \\
\label{theta2}
  \vartheta_{2n+1}(\lambda)&= \lim_{\alpha\rightarrow\infty} \frac{1}{1+e^{\alpha d_{2n+1}(\lambda)}} = \Theta_{\mathrm H}(|\lambda|-\pi/4),
\end{align}
where $\Theta_{\mathrm H}(x)$ is the Heaviside step function. 
The associated total densities are obtained using 
the BGT equations as 
\begin{align}
  \rho_{\mathrm t, 1}(\lambda)&=s(\lambda), \\
  \rho_{\mathrm t, 2}(\lambda)&=[s \star (s \cdot \Theta_{\mathrm H}(|x|-\pi/4))](\lambda) \\ \nonumber
  & \quad = \frac{1}{4 \pi^{2}}\, \sum_{k \in \mathbb Z} \frac{e^{-2 i k \lambda}}{\cosh(k\eta)}
  \left( \sum_{\ell \ne k} \frac{ \sin( (k-\ell) \pi) - \sin ((k - \ell)\pi/2)}{(k-\ell) 
  \cosh(\ell \eta)} + \frac{\pi}{2 \cosh(k \eta)} \right), \\
  \label{ccc}
  \rho_{\mathrm t, n}(\lambda)&=0, \qquad (n>2).
\end{align}
These results imply that the min entropy is completely determined by the first two densities with $n=1$ 
and $n=2$, in contrast with the quench from the N\'eel state~\cite{AlCa17b}, 
where only the first density enters in the expression for $S^{(\infty)}$. 

To derive the general expression for the min entropy,  a crucial preliminary observation is that the macrostate 
 describing $S^{(\infty)}$ has zero Yang-Yang entropy. 
 This is a general result that holds for quenches from arbitrary states. 
Indeed, first we notice that the ansatz $\eta_n=e^{\alpha\gamma_n}$ implies that, in the limit $\alpha\to\infty$,
$\vartheta_n$ can be only zero or one.
Then,  assuming that $\rho_{\textrm{t},n}$ is finite,   the Yang-Yang entropy  
\begin{equation}
S_{YY}=-L\sum_n\int d\lambda \rho_{\textrm{t},n}[\vartheta_{n}\ln\vartheta_n+
(1-\vartheta_n)\ln(1-\vartheta_{n})],
\end{equation}
must vanish in the limit $\alpha\to\infty$. 
Consequently the $ S^{(\infty)}_{\textrm{GGE}}$ is determined only by the driving functions as 
\begin{equation}
S^{(\infty)}_{\textrm{GGE}}=L\sum_n\int_{-\pi/2}^{\pi/2} d\lambda 
g_n\rho_n+L f_\textrm{GGE}. 
\label{scopy}
\end{equation}
%
Interestingly, in the large $\Delta$ limit Eq.~\eqref{scopy} simplifies. Specifically, 
only the first two strings with $n=1,2$ contribute in~\eqref{scopy}, as it is 
clear from~\eqref{ccc}. 

Upon lowering $\Delta$, we observe a sharp transition in the behaviour of $S^{(\infty)}$.
Indeed,  there is a ``critical'' value $\Delta^*$, such that for $\Delta<\Delta^*$, higher-order strings become important.
The condition that determines $\Delta^*$ is that $\gamma_2$ becomes positive, i.e., that for some $\lambda$ 
\begin{equation}
d_2+2s\star d_1\ge0.
\label{cond}
\end{equation}
The value of $\Delta^{*}$ can be found by numerically 
by imposing equality in \eqref{cond} and the final result is $\Delta^{*}\approx 1.7669$. 
This is the same value of $\Delta^*$ found for the quench from the N\'eel state~\cite{AlCa17b}, although 
the condition for higher strings to contribute for the N\'eel state (i.e. $d_1+2s\star d_2\ge0$) may appear different.
This, however, is equivalent to~\eqref{cond} after noticing 
that $d_{2n}^{\textrm{Dimer}}=d_{2n+1}^{\textrm{Neel}}$.

Finally, in contrast with the large $\Delta$ limit, for 
$\Delta<\Delta^{*}$, an analytical 
solution of \eqref{eq:gSaddlePointEquationsAlphaInfinity} 
is not possible. However, the system~\eqref{eq:gSaddlePointEquationsAlphaInfinity} 
can be effectively solved numerically. The result for $S^{(\infty)}$ 
is reported in Figure~\ref{fig:dimerRenyiEntropies} (bottom line).

\section{R\'enyi entropies of generic thermodynamic macrostates}
\label{sec:sourceTermExtraction}

In this section we show how to generalise the approach of Ref.~\cite{AlCa17} for the   
calculation of R\'enyi entropies in the case when the overlaps of a given initial  state are not known.
In this case, we just know the rapidity densities of the macrostate, e.g. from the generalised Gibbs ensemble 
construction \cite{ilievski-2015a,pvc-16}.
The key idea is embarassing simple: 
using the TBA equations (cf. ~\eqref{eq:eRenyiEntropySaddlePointEquationsCoupled} for $\alpha=1$)
\begin{equation}
\ln  \eta_{n} =  g_{n} + \sum_{m=1}^{\infty} A_{nm} 
\star \ln  [1 +1/\eta_{m}],
\label{tba2}
\end{equation}
we can extract the numerical values of  the driving functions $g_n(\lambda)$ from the saddle point solutions 
$\eta_n(\lambda)$.
Once the driving functions are numerically known, it is straightforward to use them in the formalism of 
Ref.~\cite{AlCa17} to obtain the steady-state R\'enyi entropies, as explained in the previous sections. 
Notice that this procedure does not only apply to stationary states after a quench, but can be used for {\it generic} 
Bethe states with arbitrary root densities, independently of where they come from. 
Furthermore, for quench problems, this procedure can be used to reconstruct the extensive part of the overlaps and hence 
to help in conjecturing the entire overlap function at finite size. 

To illustrate the validity of the approach, in the following subsections we provide exact results for the 
R\'enyi entropies after the quench from the tilted N\'eel state in the XXZ chain. 
For this family of quenches, the thermodynamic macrostates describing the post-quench 
steady states have been calculated in Ref. \cite{pvc-16} from the GGE. 
Only very recently (and after most of this paper was completed) the overlaps of these states with the Bethe ones
have been conjectured in Ref.~\cite{Pozsgay18}. 
Consequently, the results presented in the following, not only are a physical relevant application of these ideas but also 
provide a further  confirmation about the validity of the conjecture itself. 
Finally, we mention that if these ideas would have been developed earlier, they could have speed up the formulation 
of the conjecture in~\cite{Pozsgay18}.


\subsection{Quench from the tilted N\'eel: Extracting the driving}
\label{tneel:driv}

Here we  numerically extract the driving functions $g_n(\lambda)$ for the quench from the tilted N\'eel state. 
The key ingredients are the rapidity densities 
describing the post-quench steady state that have been determined in Ref.~\cite{pvc-16}. 
These are used in the system~\eqref{tba2} to extract $g_n$. 
To make the paper self contained, we report the results for the saddle point densities. 
The starting point is  
$\eta_1(\lambda)$, which is given as 
\begin{equation}
\label{eta1}
  \eta_{1}(\lambda) = -1 + \frac{ T_{1}\!\left(\lambda + i \frac{ \eta}{2 }\right) T_{1}\!\left(\lambda - i \frac{ \eta}{2 }\right)}{\phi\!\left( \lambda + i \frac{ \eta}{ 2} \right) \bar \phi\!\left( \lambda - i \frac{\eta}{2} \right)}, 
  \end{equation}
where the auxiliary functions $\phi,\bar\phi$ and $T_1$ are defined as  
 \begin{align}
  T_{1}(\lambda) &= - \frac{1}{8} \cot(\lambda) \{ 8 \cosh(\eta) \sin^{2}(\theta) \sin^{2}(\lambda) - 4 \cosh(2\eta) + [\cos(2\theta) + 3][2 \cos(2\lambda) - 1] + 2 \sin^{2}(\theta) \cos(4\lambda) \} ,  
  \label{eq:eSaddlePointTiltedNeelFunctionT1}   \\
  \phi(\lambda) &= \frac{1}{8} \sin(2\lambda + i \eta) [2 \sin^{2}(\theta) \cos(2\lambda - i \eta) + \cos(2\theta) + 3],
  \label{eq:eSaddlePointTiltedNeelFunctionPhi} \\
  \bar \phi(\lambda) &= \frac{1}{8} \sin(2\lambda - i \eta) [2 \sin^{2}(\theta) \cos(2\lambda + i \eta) + \cos(2\theta) + 3].
  \label{eq:eSaddlePointTiltedNeelFunctionPhiBar}
\end{align}
Here $\theta$ denotes the tilting angle. 
For $n>1$, $\eta_{n}(\lambda)$ is determined recursively from the Y-system
\begin{equation}
  \eta_{n+1}(\lambda) = 
  \frac{ \eta_{n}(\lambda + i \eta /2) \eta_{n}(\lambda - i \eta/2)}{ 1 + \eta_{n-1}(\lambda)}-1, 
  \label{eq:YRelations}
\end{equation}
with the convention $\eta_{0}\equiv 0$. From the densities $\eta_n$, the particle 
densities $\rho_n$ are obtained, as usual, by solving the thermodynamic 
version of the TBA equations~\eqref{eq:rBetheGaudinTakahashiCoupled}. 

%
\begin{figure}[t]
  \centering
  \includegraphics{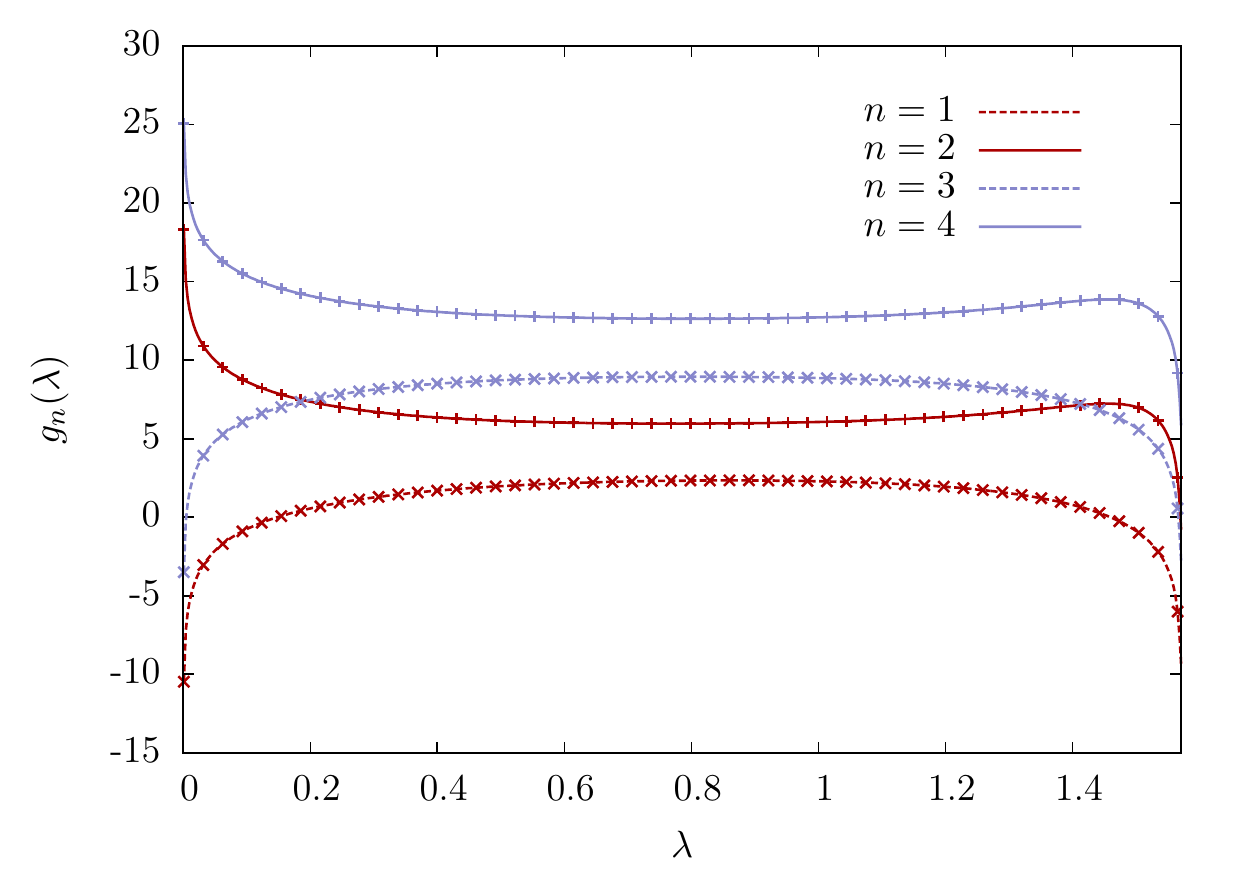}
  \caption{GGE driving functions $g_{n}$ for the quench from 
  the tilted N\'eel state in the XXZ chain with $\Delta=2$. 
  On the $x$-axis $\lambda$ is the rapidity variable. 
  The N\'eel tilting angle is $\theta=\pi/3$. Different lines 
  correspond to different strings lengths $n$. Lines are numerical 
  results using the TBA approach, whereas points denote  
  the analytical conjecture in Ref.~\cite{Pozsgay18}.}
  \label{fig:comparisonTiltedNeel}
\end{figure}
%

The driving function $g_n$ can be easily extracted numerically by plugging the above root densities in Eq. \eqref{tba2}.
The results for quenches for the XXZ chain with $\Delta=2$ and the quench from the tilted N\'eel state with 
tilting angle $\theta=\pi/3$ are reported in Figure~\ref{fig:comparisonTiltedNeel}.
These results may be compared with the  recently conjectured form of the overlaps~\cite{Pozsgay18}.. 
%
%
So far, this conjecture has been tested numerically for Bethe states containing few 
particles, and it has been shown to give the correct thermodynamic macrostate 
after the quench. The thermodynamic limit of the overlaps in~\cite{Pozsgay18} can be written as
\begin{equation}
  \ln  \langle \Psi_{0} |\rho_n\rangle = L\sum_{n=1}^{\infty} \int_{-\pi/2}^{\pi/2} \mathrm{d} \lambda \rho_{n}(\lambda) 
  g_{n}(\lambda), 
  \label{eq:pTiltedNeelOverlaps} 
 \end{equation}
where $\rho_n$ are the particle densities describing the thermodynamic macrostate and 
the explicit forms of $g_n$ read~\cite{Pozsgay18} 
 \begin{align}
  g_{1}(\lambda) &= \frac{\tan(\lambda + i \eta / 2) \tan(\lambda - i \eta / 2)}{4 \sin^{2}(2\lambda)}\cdot \frac{\cos^{2}(\lambda+i \xi)\cos^{2}(\lambda-i \xi)}{\cosh^{4}(\xi)}, 
  \label{eq:pTiltedNeelOverlapKernel1} \\
  g_{n}(\lambda) &= \sum_{j=1}^{n} g_{1}(\lambda + i \eta / 2 (n+1-2j)), 
  \label{eq:pTiltedNeelOverlapKernelN}
\end{align}
where $\xi$ is related to the tilting angle $\theta$ as  $\xi\equiv -\ln(\tan(\theta / 2))$. 

We now compare the driving functions $g_n$ as extracted from 
the TBA equations~\eqref{tba2} 
with the conjectured result in Eq.~\eqref{eq:pTiltedNeelOverlapKernel1}
and \eqref{eq:pTiltedNeelOverlapKernelN}. 
The comparison is presented in Figure~\ref{fig:comparisonTiltedNeel} for the 
XXZ chain with $\Delta=2$ and the quench from the tilted N\'eel state with 
tilting angle $\theta=\pi/3$ (we tested them also for other tilting angles, finding equivalent results 
that we do not report here). The continuous lines are the numerical results 
for $g_n$ for $n\le 4$ (higher strings are not reported). The different symbols (crosses) are 
the numerical results using the conjecture~\eqref{eq:pTiltedNeelOverlapKernel1}-\eqref{eq:pTiltedNeelOverlapKernelN}. As it is clear from the Figure, the agreement 
between the two results is perfect for all values of $\lambda$. 

\subsection{R\'enyi entropies after quenching from the tilted N\'eel state}
\label{num-res}

%
\begin{figure}[t]
\centering
\includegraphics{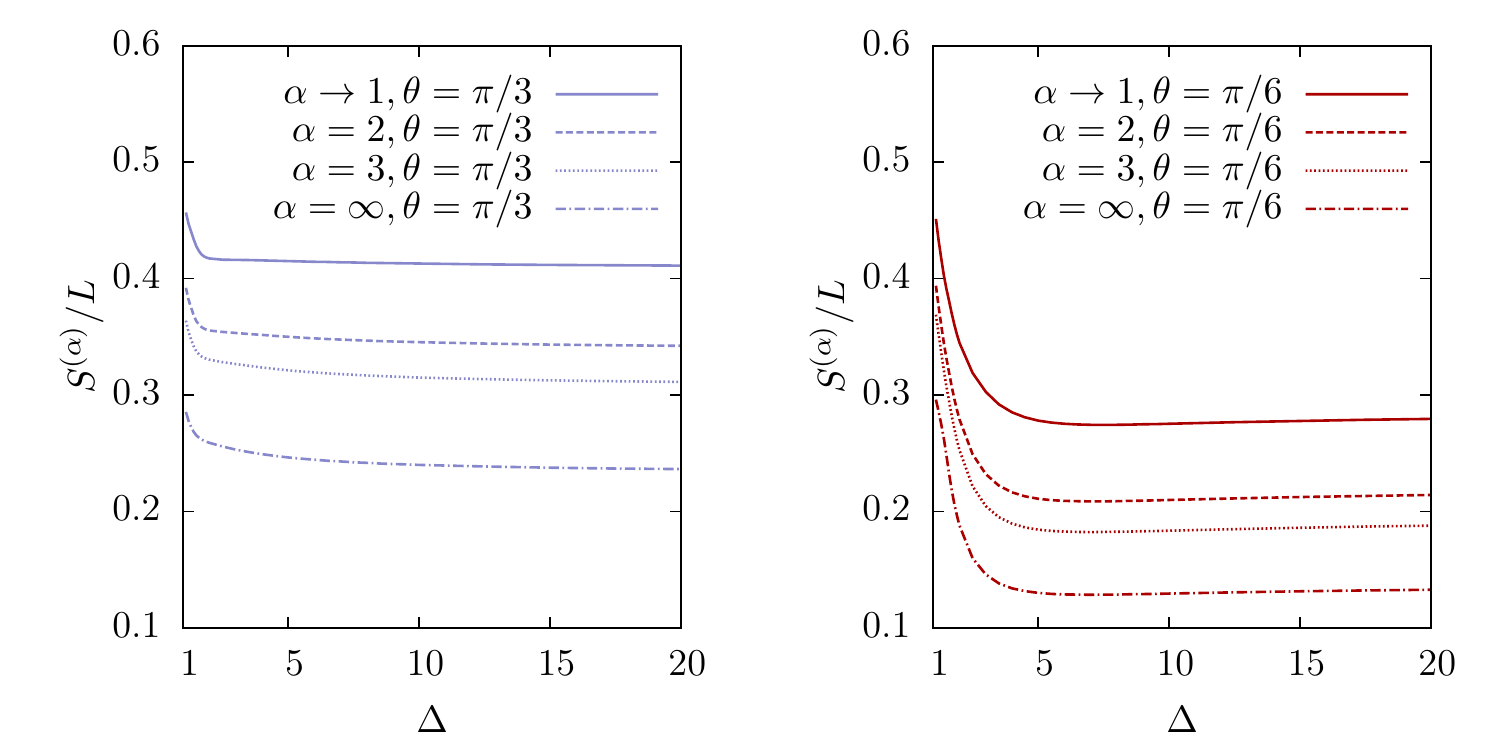}
\caption{Steady-state R\'enyi entropies $S^{(\alpha)}/L$ after the 
 quench from the tilted N\'eel state in the XXZ spin chain. The entropies 
 are plotted as a function of the anisotropy $\Delta$. We show results 
 for tilting angles $\theta=\pi/6$ and $\theta=\pi/3$ (right and 
 left panel, respectively). In each panel the different curves correspond 
 to different values of $\alpha\in[1,\infty)$.
}
\label{fig:rtheo}
\end{figure}
%

Now we are in the position to obtain results for the steady-state 
R\'enyi entropies after the quench from the tilted N\'eel state in the XXZ chain. 
The theoretical predictions for the entropies are obtained by combining 
the results of section~\ref{tneel:driv} to extract the driving functions $g_n$ 
with the procedure of Ref.~\cite{AlCa17} (see section~\ref{sec:renyiEntropyTBA}). 
Our results are reported in 
Figure~\ref{fig:rtheo} plotting $S^{(\alpha)}/L$ versus the chain anisotropy 
$\Delta$. The data shown in the Figure are for the tilted N\'eel with 
tilting angle $\theta=\pi/3$ and $\theta=\pi/6$ (right and left panel, 
respectively). As expected, one has that $S^{(\alpha)}<S^{(\alpha')}$ 
if $\alpha>\alpha'$. 
%
%
For all values of $\alpha$ and $\theta\neq0$ the entropy densities 
are finite in the limit $\Delta\to\infty$, in contrast with the quench from 
the N\'eel state~\cite{AlCa17b}, where all the entropies vanish for $\Delta\to\infty$. 
Finally, an intriguing feature is that for $\theta=\pi/6$ the behaviour of the 
entropies is not monotonic as a function of $\Delta$, but $S^{(\alpha)}$ exhibits 
a minimum around $\Delta\approx 5$, although the minimum is not very pronounced. 
This remains true for a window of tilting angle $\theta$ close to $\pi/6$.

\subsection{The min entropy}
\label{sc-phase}

%
\begin{figure}[t]
\centering
\includegraphics{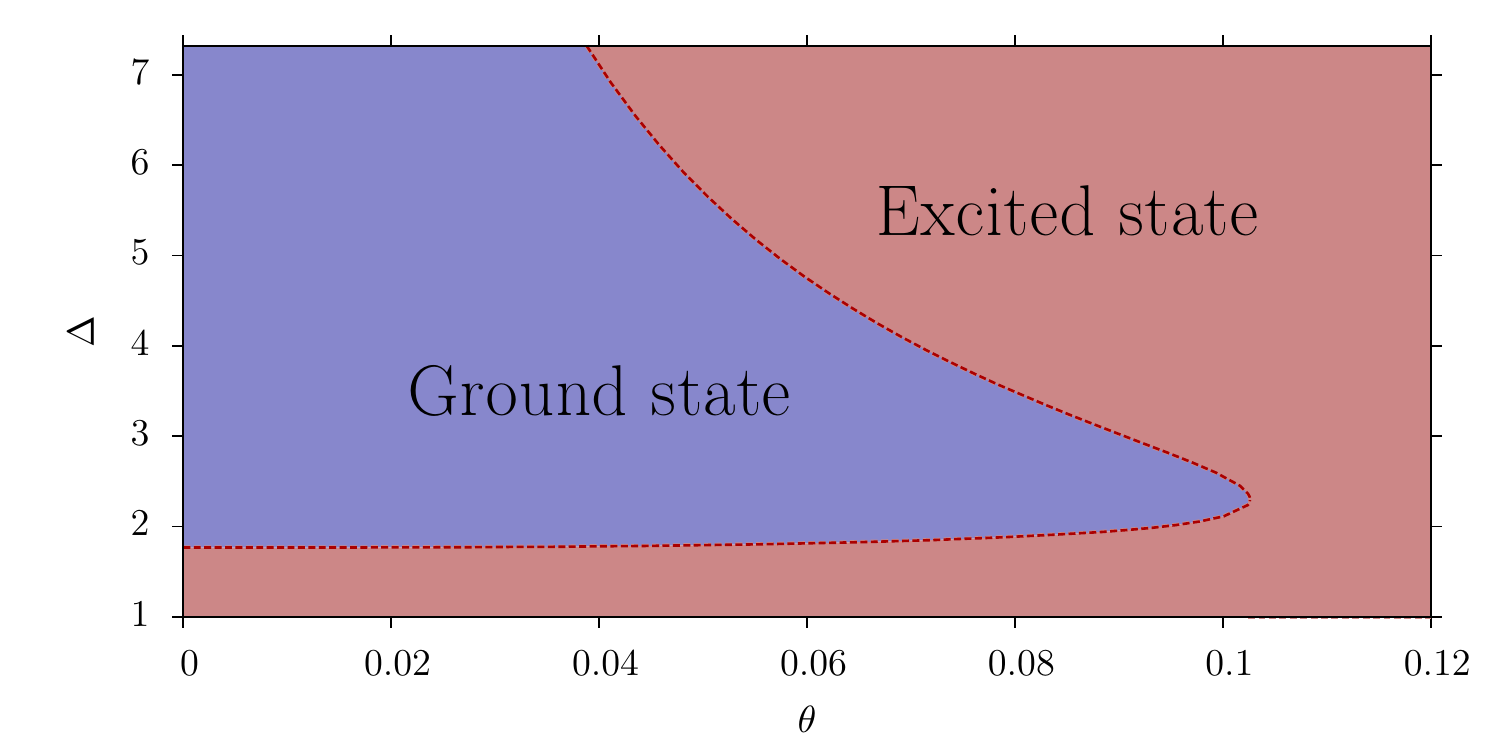}
\caption{TBA macrostate for the steady-state value of the 
 min entropy $S^{(\infty)}$ after the quench 
 from the tilted N\'eel state in the XXZ chain. The Figure show 
 the macrostate as a function of the chain anisotropy $\Delta$ 
 and the tilting angle $\theta$. The blue area denotes the 
 parameter region $(\theta,\Delta)$ where the macrostate 
 is the ground state of the XXZ chain at that $\Delta$. 
 Outside this region, the macrostate is an excited state of the 
 chain. The dashed line divides the two regions. 
}
\label{fig:phase}
\end{figure}
%

We now focus on the steady-state value of the min entropy. Similar to 
the N\'eel and dimer states, in the limit $\alpha\to\infty$ one can use the ansatz 
$(\ln\eta_n)/\alpha=\gamma_n$. The equations for $\gamma_n$ are the 
same as for the dimer (cf.~\eqref{eq:gSaddlePointEquationsAlphaInfinity}), 
i.e., 
\begin{equation}
\gamma_n=d_n+s\star(\gamma_{n-1}^++\gamma_{n+1}^+), 
\end{equation}
where now the driving $d_n$ is obtained from the 
driving functions $g_n$ for the quench from the tilted N\'eel 
as 
\begin{equation}
d_n=g_n-s\star(g_{n+1}+g_{n-1}). 
\end{equation}
For the N\'eel quench, i.e., for $\theta=0$, there is 
a ``critical'' value $\Delta^*$ such that for $\Delta>\Delta^*$ the 
thermodynamic macrostate that the describes the min entropy
is the ground state of the XXZ chain~\cite{AlCa17b}. For $\Delta<\Delta^*$ this is not 
the case, and the macrostate is an excited state with zero Yang-Yang entropy. 
The ``critical'' $\Delta^*$ at which the behaviour of the thermodynamic macrostate changes is determined 
for the N\'eel state (and for the dimer state as well) 
by the condition 
that $\gamma_2$ becomes positive.  It is natural to investigate how this 
scenario is modified  upon tilting the initial state. 
Here we show that the macrostate describing the min entropy is the ground state of the XXZ chain provided 
that the tilting angle $\theta$ is not too large. 

To clarify this issue, we numerically observed that for large $\Delta$  
\begin{equation}
  \begin{split}
\label{cond1}
& \gamma_1(\lambda)<0,\\
& \gamma_3(\lambda)<0. 
\end{split}
\end{equation}
The conditions in Eq.~\eqref{cond1} have important consequences for the 
particle densities $\rho_n$. In particular, it implies that the macrostate describing 
the min entropy is the ground state of the XXZ chain. 
To show that, let us consider the TBA equations for $\rho_{\textrm{t},n}$  
\begin{equation}
\rho_{n,\textrm{t}}=s\star[(1-\vartheta_{n-1})\rho_{n-1,\textrm{t}}+
(1-\vartheta_{n+1})\rho_{n+1,\textrm{t}}]. 
\label{sys}
\end{equation}
First, since $\vartheta_0=0$ and $\rho_0=\delta(\lambda)$, one has that 
$\rho_1=s$, which is the density of the ground state of the XXZ chain. 
Clearly, the conditions in~\eqref{cond1} together with 
the system~\eqref{sys} imply that $\rho_{2,\textrm{t}}=0$. Another 
important consequence is that 
the first two equations in~\eqref{sys} are decoupled from the rest, 
which form a linear homogeneous system of integral 
equations. Moreover, for $n\to\infty$ one expects that $\rho_{\textrm{t},n}
\to 0$. Thus, it is natural to conjecture that $\rho_{\textrm{t},n}=0$ 
for any $n>2$. 
Finally, we observe that a similar decoupling occurs for the 
quench from the N\'eel state~\cite{AlCa17b}, although via a different mechanism. Precisely, 
for the the N\'eel state one has that $\gamma_{2n+1}<0$ for all $n$. 

We now use the  conditions~\eqref{cond1}  to characterise the behaviour of the min entropy. 
Our results are summarised in the ``phase diagram'' in Fig.~\ref{fig:phase}. 
The blue region in the figure corresponds to the region in the parameter space $(\theta,\Delta)$ 
where the thermodynamic macrostate describing the min entropy is the ground state of the XXZ chain (at that value of $\Delta$). 
For $\theta=0$ we recover the result of Ref.~\cite{AlCa17b}, i.e., that the ground state describes the min entropy
for $\Delta>\Delta^*\approx 1.766$. 
The ground state remains the correct macrostate for the min entropy in a region of not too large $\theta$. 
Conversely, for $\theta\gtrsim 0.1$ the macrostate is an excited state of the XXZ at any $\Delta$. 
However, at smaller $\theta$ there is always an extended region where the min entropy
is described by the ground state of the XXZ chain. 
The extension of this region shrinks upon increasing $\Delta$, and it is likely to vanish in the limit $\Delta\to\infty$. 
The dashed line in Fig.~\ref{fig:phase} marks the ``transition'' between the two regimes. 
The line is obtained by numerically finding the values of $(\theta,\Delta)$ for which either $\gamma_1$ or $\gamma_3$ 
vanish, violating the conditions in Eq.~\eqref{cond1}.

\section{Numerical benchmarks using tDMRG}
\label{sec:dmrg}

%
\begin{figure}[t]
\centering
\includegraphics{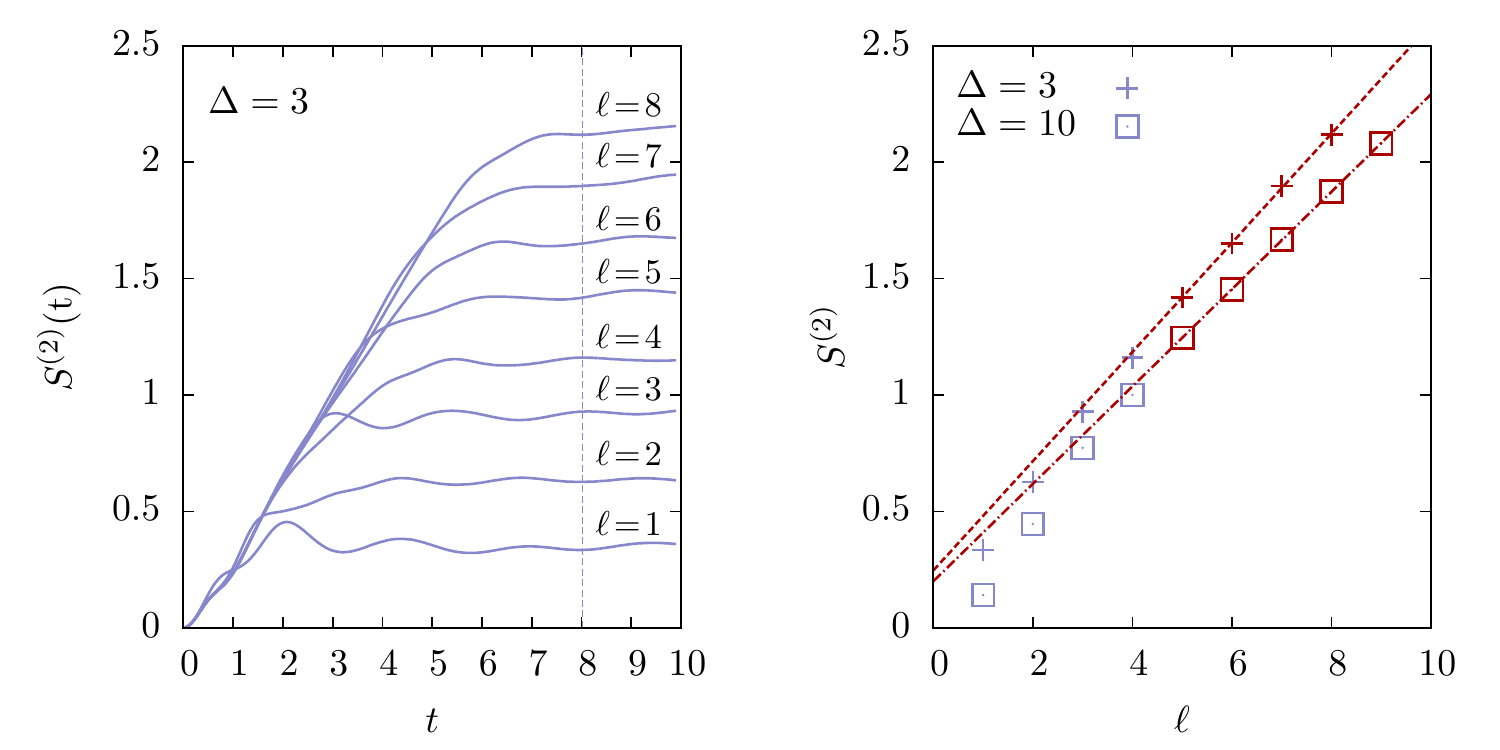}
\caption{Testing TBA results for the steady-state R\'enyi entropies 
 against tDMRG simulations in the XXZ chain. The system is quenched 
 from the tilted N\'eel state with different values of the chain 
 anisotropy $\Delta$ and tilting angle $\theta$. Left panel: tDMRG 
 data for the second R\'enyi entanglement entropy $S^{(2)}$ 
 at $\Delta=3$ plotted as a function of time. Different lines are 
 for different sizes $\ell$ of the subsystem. 
 Right panel: Saturation values of the entropy density $S^{(2)}$ at long time (i.e. data 
 at $t=8$, corresponding to the vertical line in the left panel). The entropy 
 density (symbols in the figure) is plotted versus $\ell$. The 
 lines are linear fits to $a\ell+b$. Only the red symbols are 
 included in the fit. The fit gives 
 $a\approx0.23$ and $a\approx0.21$ for $\Delta=10$ and $\Delta=3$, 
 respectively. These values are compatible with the TBA 
 results $S^{(2)}/\ell \approx 0.2321$ for $\Delta=3$ and 
 $S^{(2)}/\ell \approx 0.2098$ for $\Delta=10$.
}
\label{fig:dmrg}
\end{figure}
%

In this section we exploit the equivalence between the reduced density matrices in the thermodynamic ensemble (the GGE in our case)
and in the long time limit after a quench to provide numerical confirmations of the results of the previous section by means of 
entanglement entropy dynamics. 
We employ tDRMG~\cite{white-2004,daley-2004,uli-2005,uli-2011} simulations in the framework of the Matrix Product States (MPS) 
\cite{itensor}. 
The tilted N\'eel state is conveniently constructed by first constructing 
the N\'eel state, which admits a simple MPS representation with 
bond dimension $\chi=1$. The tilted N\'eel is then obtained 
by applying a global rotation site by site, by using the Matrix Product 
Operator representation of the spin rotation operator 
$\exp(i\theta\sum_i\sigma_i^y)$. The time evolution is implemented 
by using a standard second order Trotter-Suzuki approximation of 
the time evolution operator $\exp(-i H t)$, with a time step 
$\delta t=0.02$ . At each application of the time evolution the 
bond dimension $\chi$ grows. To keep $\chi$ 
reasonably small, at each time step we perform a truncated 
Singular Value Decomposition (SVD) of each tensor forming the MPS. 
Specifically, in the SVD we keep the largest $\chi_{\textrm{max}}$ 
singular values, with $\chi_{\textrm{max}}\approx 100$. For the 
quenches that we consider we verified that this is sufficient to 
obtain accurate results up to times $t\approx 10$. 

Our numerical results are shown in Figure~\ref{fig:dmrg}. The left 
panel shows the second R\'enyi entropy  $S^{(2)}$ plotted as  a function of time. The data are tDMRG results 
for the quench from the tilted N\'eel state in the XXZ chain for 
tilting angle $\theta=\pi/6$ and $\Delta=3$. The different lines 
correspond to different subsystem 
sizes $\ell$. The right panel shows the saturation value of 
$S^{(2)}$ as a function of $\ell$. Precisely, for $\Delta=3$ the 
data correspond to $t\approx 8$ (see vertical line in the left 
panel). The different symbols correspond to different values of 
$\Delta$. The expected volume-law behaviour $S^{(2)}\propto\ell$ 
is clearly visible. The lines are linear fits to $a\ell+b$, with $a,b$ 
fitting parameters. Only the largest sizes $\ell$ are included 
in the fit (red symbols in the Figure). The fit gives $a\approx0.23$ 
and $a\approx 0.21$ 
for $\Delta=3$ and $\Delta=10$, respectively. These results are in 
excellent agreement with the TBA predictions $a\approx0.2321$ 
and $a=0.2098$ (see Figure~\ref{fig:rtheo} for the predictions). 

\section{Conclusions}
\label{sec:conclusions}

In this manuscript we studied the R\'enyi entropies in the stationary state after a quantum quench. 
As shown in Refs.~\cite{AlCa17,AlCa17b}, in the  quench action approach the R\'enyi entropies are generalised free energy 
of a macrostate that may be derived from the knowledge of the overlaps of the initial state with Bethe eigenstates.
The thermodynamic limit of the overlaps provides the driving term in the TBA formalism. 
Here we considered the problem of determining the R\'enyi entropies in a generic thermodynamic macrostate of integrable models, 
even in those cases when the overlaps are not known. 
We showed that the needed driving term can be reconstructed starting from the macrostate's particle densities.
Then we provided a major simplification of the expression for the generalised free energy that may be rewritten only as a function
of the occupation numbers of one-strings, cf. Eq. \eqref{eq:sRenyiEntropyTheta1Expression} which is a much simpler and 
manageable formula than the known sum over all string content. 

We then studied accurately the stationary R\'enyi entropies after the quench from the dimer and the tilted N\'eel states in the 
XXZ Heisenberg spin chain.
For the former initial state we employed the overlap TBA approach, while for the latter we reconstructed the driving terms from the macrostate.
The overall results for the dimer states are summarised in Fig. \ref{fig:dimerRenyiEntropies} which shows the $\Delta$ 
and $\alpha$ dependence of the R\'enyi entropies.
We also analysed in details two limits that are analytically tractable, namely $\Delta\to\infty$ and $\alpha\to \infty$.
In the Ising limit $\Delta\to\infty$, the result for the R\'enyi entropies resembles that of free-fermion models, as it should. 
Deviations from the free-fermion result appear already at the first non trivial order in $1/\Delta$. 
For the min entropy, i.e. $\alpha\to \infty$, we found that the representative state has vanishing Yang-Yang entropy for arbitrary 
$\Delta$.
We also found a sharp transition of this state at a critical value of $\Delta$ denoted as $\Delta^*$. 
For $\Delta>\Delta^*$, the representative state contains one- and two-strings only (as a difference with the quench 
from N\'eel state where only one-strings matter) while for $\Delta<\Delta^*$ the other bound states start being present.
When the initial configuration is the tilted N\'eel state, the results for the R\'enyi entropies as function of $\Delta,\alpha$, and 
the tilting angle $\theta$ are reported in Fig. \ref{fig:rtheo}. As a main difference with the other cases, the entropies as 
function of $\Delta$ are not always monotonic, but they may show a minimum for some values of $\theta$. 
Also in this case we analytically studied the min entropy.
We again found that the representative eigenstate has zero Yang-Yang entropy for arbitrary $\Delta$ and $\theta$ and that 
there is a sharp transition line. 
The resulting "phase diagram" is reported in Fig. \ref{fig:phase}: there is a region for small $\theta$ where 
the representative eigenstate is the ground state (and hence only one-strings are present), while in the rest of the 
phase diagram, other bound states matter. 
The results presented here (and the ones for the N\'eel state \cite{AlCa17b}) show that rather generically the 
representative state of the min entropy has zero Yang-Yang entropy. 
It would be interesting to find out the minimal conditions on the initial state for this property to be generically valid. 

Finally, we exploited the equivalence between thermodynamic and entanglement entropies to check by means 
of numerical simulations the correctness of our results. 
We found that the numerical data for the R\'enyi entanglement entropies at large time are perfectly compatible with TBA results
for the thermodynamic entropies. 

A major problem that remains still open is to characterise the time evolution of R\'enyi entanglement entropies for generic 
interacting integrable model, both for homogeneous quenches and quenches from piecewise homogeneous initial states (see \cite{Alba18} for some results). 
Technically, it is not possible to generalise the semiclassical  approach developed for the von Neumann entropy \cite{alba-2016,alba-2018}
because the R\'enyi entropies have been written in terms of root distributions which are not the ones 
of the macrostate describing local properties: only the latter determines the asymptotic spreading of 
entanglement \cite{alba-2016,alba-2018} and correlations \cite{bonnes-2014}. 
Apart from the {\it per se} theoretical interest, this issue is also fundamental for a comparison with cold atom experiments 
in which only R\'enyi entropies can be measured \cite{islam-2015,daley-2012,evdcz-18,kaufman-2016}.


\section*{Acknowledgments}
M.M. thanks Bal\'azs Pozsgay for inspiring discussions. V.A. acknowledges support from the European Union’s Horizon 2020 under the Marie Sklodowoska-Curie grant agreement No 702612 OEMBS. Part of this work has been carried out during the workshop ``Quantum paths''  at  the  Erwin  Schrödinger  International  Institute for  Mathematics  and  Physics (ESI) in Vienna, and during the workshop ``Entanglement in Quantum Systems'' at the Galileo Galilei Institute (GGI) in Florence.

\end{document}